\PassOptionsToPackage{table,dvipsnames}{xcolor}
\documentclass[sigconf]{acmart}
\AtBeginDocument{%
  }

\setcopyright{acmlicensed}
\copyrightyear{2026}
\acmYear{2026}
\acmDOI{XXXXXXX.XXXXXXX}
\acmConference[Conference acronym 'XX]{Make sure to enter the correct
  conference title from your rights confirmation email}{June 03--05,
  2026}{Woodstock, NY}
\acmISBN{978-1-4503-XXXX-X/2018/06}

\hypersetup{
  colorlinks=true,
  linkcolor=blue,
  citecolor=blue,
  urlcolor=blue
}
\usepackage{fancyvrb}
\usepackage{amsopn}

\usepackage{subcaption}

\usepackage{microtype,xspace,graphicx,multirow}
\usepackage{booktabs}
\usepackage{array,underscore,relsize}
\usepackage{enumitem}
\usepackage{adjustbox}
\usepackage{tabularx,longtable}
\usepackage{rotating,pdflscape}
\usepackage{makecell}
\usepackage{pifont}
\usepackage{fp}
\usepackage{wasysym}
\usepackage[most]{tcolorbox}
\usepackage[linesnumbered,ruled,vlined]{algorithm2e}
\usepackage{minted}
\usepackage[normalem]{ulem}

\usepackage{siunitx}
\usepackage{tikz}
\usetikzlibrary{patterns}

\usepackage{pgfplots}
\usepgfplotslibrary{statistics}
\usepackage{kotex}

\newcommand{\cc}[1]{\mbox{\smaller[0.5]\texttt{#1}}}

\fvset{fontsize=\scriptsize,xleftmargin=8pt,numbers=left,numbersep=5pt}

\def\Snospace~{\S{}}

\newif\ifdraft\drafttrue
\newif\ifnotes\notestrue
\ifdraft\else\notesfalse\fi

\input{glyphtounicode}
\newcolumntype{R}[1]{>{\raggedleft\let\newline\\\arraybackslash\hspace{0pt}}p{#1}}

\newcommand{\squishlist}{
\begin{itemize}[noitemsep,nolistsep]
  \setlength{\itemsep}{-0pt}
}
\newcommand{\squishend}{
  \end{itemize}
}

\colorlet{bluekeywords}{blue!90!black}
\colorlet{greencomments}{green!40!black}
\colorlet{redstrings}{red!90!black}
\newcommand*{\algokeywordsty}[1]{\textcolor{bluekeywords}{\textbf{\ttfamily{#1}}}}

\newcommand*{\Not}{\algokeywordsty{not}~}

\usepackage{tikz}
\newcommand*\WC[1]{%
\begin{tikzpicture}[baseline=(C.base)]
\node[draw,circle,inner sep=0.2pt](C) {#1};
\end{tikzpicture}}

\usepackage{xstring}
\newcommand{\PP}[1]{
\vspace{2px}
\noindent{\bf \IfEndWith{#1}{.}{#1}{#1.}}
}

\definecolor{promptbg}{gray}{0.97}
\definecolor{promptframe}{gray}{0.75}
\definecolor{xmltag}{RGB}{0,102,153}
\definecolor{xmlcontent}{RGB}{51,51,51}

\newtcolorbox{promptbox}[2][]{%
  colback=promptbg,
  colframe=promptframe,
  fonttitle=\bfseries\small,
  coltitle=black,
  title=#2,
  boxrule=0.5pt,
  left=2pt,
  right=2pt,
  top=2pt,
  bottom=2pt,
  width=\columnwidth,
  before={\par\medskip\noindent},
  after={\par\medskip},
  enhanced,
  arc=3pt,
  #1
}
\newtcolorbox{promptboxwrap}[2][]{%
  colback=promptbg,
  colframe=promptframe,
  fonttitle=\bfseries\small,
  title=#2,
  boxrule=0.5pt,
  left=2pt,
  right=2pt,
  top=2pt,
  bottom=2pt,
  width=\textwidth,
  before={\par\medskip\noindent},
  after={\par\medskip},
  listing only,
  listing options={
    language=XML,
    basicstyle=\ttfamily\footnotesize,
    breaklines=true,
    breakatwhitespace=true,
    columns=flexible,
    showspaces=false,
    showstringspaces=false,
    frame=none,
    numbers=none,
    xleftmargin=0pt,
    xrightmargin=0pt
  },
  #1
}
\newtcolorbox{promptboxinline}[2][]{%
  colback=promptbg,
  colframe=promptframe,
  fonttitle=\bfseries\small,
  coltitle=black,
  title=#2,
  boxrule=0.5pt,
  left=2pt,
  right=2pt,
  top=2pt,
  bottom=2pt,
  breakable,
  width=\columnwidth,
  enhanced,
  arc=3pt,
  #1
}

\newcommand{\etal}{\textit{et al}.\xspace}

\newcommand{\eg}{e.g.} 

\newcommand{\coordinator}{\textsc{Coordinator}\xspace}
\newcommand{\worker}{\textsc{Worker}\xspace}
\newcommand{\workers}{\textsc{Workers}\xspace}

\newcommand{\sys}{{\textsc{PatchIsland}}\xspace}

\newcommand{\syscrete}{\mbox{\textsc{Crete}}\xspace}
\newcommand{\sysmartian}{\mbox{\textsc{Martian}}\xspace}
\newcommand{\sysmr}{\mbox{\textsc{MultiRetrieval}}\xspace}
\newcommand{\sysprism}{\mbox{\textsc{Prism}}\xspace}
\newcommand{\sysvincent}{\mbox{\textsc{Vincent}}\xspace}
\newcommand{\sysclaudelike}{\mbox{\textsc{ClaudeLike}}\xspace}
\newcommand{\sysaider}{\mbox{\textsc{Aider}}\xspace}
\newcommand{\syssweagent}{\mbox{\textsc{SWE-Agent}}\xspace}

\newcommand{\fp}{FP$^2$\xspace}

\newcommand{\baselinea}{\mbox{\textsc{RoboDuck}}\xspace}
\newcommand{\baselineb}{\mbox{\textsc{Buttercup}}\xspace}
\newcommand{\teama}{\mbox{\textsc{Theori}}\xspace}
\newcommand{\teamb}{\mbox{\textsc{Trail of Bits}}\xspace}
\newcommand{\teamc}{\mbox{\textsc{Lacrosse}}\xspace}
\newcommand{\teamd}{\mbox{\textsc{ALL YOU NEED IS A FUZZING BRAIN}}\xspace}
\newcommand{\teame}{\mbox{\textsc{Shellphish}}\xspace}
\newcommand{\teamf}{\mbox{\textsc{42-b3yond-6ug}}\xspace}

\newcommand{\SOUND}{\textcolor{green}{\checkmark}}
\newcommand{\BAD}{\textcolor{red}{\ding{55}}}
\newcommand{\PASS}{\textsf{\textbf{\textcolor{green!60!black}{\checkmark\ pass}}}}
\newcommand{\FAIL}{\textsf{\textbf{\textcolor{red!70!black}{\texttimes\ fail}}}}

\newcolumntype{C}[1]{>{\centering\arraybackslash}m{#1}}
\newcolumntype{L}[1]{>{\raggedright\arraybackslash}m{#1}}
\newcommand{\n}{\\}

\usepackage{fp}
\newcommand{\ratio}[2]{%
  \FPeval{\temp}{round(100*#1/#2,1)}%
  \temp\%%
}
\title{\sys: Orchestration of LLM Agents for Continuous Vulnerability Repair}

\ifdefined\DRAFT
\fi

\author{
\large
Wonyoung Kim$^{1,6,*}$,
Seunggi Min$^{1,*}$,
Minjae Gwon$^{2}$,
Dowoo Baik$^{5}$,
Haein Lee$^{1}$,
Hyeon Heo$^{7}$,
Minjae Lee$^{2}$,
Min Woo Baek$^{1}$,
Yonghwi Jin$^{3}$,
Younggi Park$^{4}$,
Yunjae Choi$^{6}$,
Taesoo Kim$^{3,8,\dagger}$,
Sangdon Park$^{2,\dagger}$,
Insu Yun$^{1,\dagger}$
\\[2.0em]
\normalsize
$^{1}$ KAIST \quad
$^{2}$ POSTECH \quad
$^{3}$ Georgia Institute of Technology \quad
$^{4}$ Korea University \\[0.2em]
$^{5}$ Samsung SDS \quad
$^{6}$ Samsung Electronics \quad
$^{7}$ ENKI WhiteHat \quad
$^{8}$ Microsoft
\\[1.0em]
}

\date{}

\setcopyright{none}
\renewcommand\footnotetextcopyrightpermission[1]{} 
\begin{document}
\begin{abstract}
Continuous fuzzing platforms such as OSS-Fuzz uncover large numbers of vulnerabilities, yet the subsequent repair process remains largely manual. Unfortunately, existing Automated Vulnerability Repair (AVR) techniques---including recent LLM-based systems---are not directly applicable to continuous fuzzing. This is because these systems are designed and evaluated on a static, single-run benchmark setting, making them ill-suited for the diverse, noisy, and failure-prone environments in continuous fuzzing.

To address these issues, we introduce \sys, a system for Continuous Vulnerability Repair (CVR) that tightly integrates with continuous fuzzing pipelines. \sys employs an ensemble of diverse LLM agents. By leveraging multiple LLM agents, \sys can cover a wider range of settings (e.g., different projects, bug types, and programming languages) and also improve operational robustness. In addition, \sys utilizes a two-phase patch-based deduplication to mitigate duplicate crashes and patches, which can be problematic in continuous fuzzing. 

In our internal evaluation, \sys repaired 84 of 92 vulnerabilities, demonstrating strong repair capability. In the official AIxCC competition, the system operated with no human intervention in a fully autonomous environment and successfully patched 31 out of 43 vulnerabilities, achieving a repair rate of 72.1\%.
\end{abstract}

\maketitle
\pagestyle{plain}

\ifdefined\fancyhf
  \fancyhf{}
\fi

\makeatletter
\markboth{}{}
\makeatother


\section{Introduction}
\label{s:intro}

In recent years, automated vulnerability discovery has advanced substantially, leading to the widespread integration of fuzzing into development workflows~\cite{PeachCI,EffectiveFuzzingCI}. Notably, OSS-Fuzz~\cite{ossfuzz} established a continuous fuzzing pipeline and has identified a large number of vulnerabilities across diverse open-source projects.

Despite these advances, post-fuzzing steps, including crash triage and patch generation, remain largely manual.
As the volume of discovered vulnerabilities continues to grow, these steps demand increasing resources and human effort,
becoming a major bottleneck in continuous fuzzing pipelines.
To address this challenge, Automated Vulnerability Repair (AVR) has emerged as a promising approach.
Thanks to recent advances in LLMs, LLM-based AVR techniques have demonstrated remarkable performance
~\cite{appatch, san2patch, csuvik2024genprogjs} for real-world vulnerabilities.

Unfortunately, despite their impressive performance, existing AVR techniques are not directly applicable to continuous fuzzing.
This is because these techniques are designed for a static, single-run workload.
In contrast, continuous fuzzing uncovers vulnerabilities at scale and over time,
rendering these techniques difficult to apply.
Moreover, for continuous fuzzing, AVR systems must be operationally robust;
They need to seamlessly integrate with various environments (e.g., projects, build configurations, and vulnerability types).
Unfortunately, existing AVR techniques are not designed with this robustness in mind.

To bridge this gap, we propose \sys, a continuous vulnerability repair (CVR) system for continuous fuzzing.
To address challenges in continuous fuzzing, \sys employs several novel techniques.
First, we leverage an \emph{ensemble-based approach} that exploits the diversity of multiple agents to improve operational robustness and effectiveness.
Second, \sys employs two-phase deduplication to mitigate duplicate crashes and patches, which are common in continuous fuzzing.
Finally, \sys employs \fp (First-come first-served, Preference-based, and Provider-aware) orchestration to schedule agents.
This approach achieves robustness while minimizing the overhead associated with multiple agents.
We also introduce \syscrete, a framework for building agents, to reduce repeated engineering effort and maintain consistency for building diverse agents.

We evaluated \sys on the AIxCC benchmark and in the AIxCC final competition.
Overall, \sys repaired 84 of 92 vulnerabilities, outperforming the baseline system, \baselineb, by 29.3\%.
In the AIxCC final competition, which runs over a week with 53 challenge projects,
\sys emerged as the most successful system in patch generation;
it repaired the largest number of vulnerabilities (31) among all participating systems 
and achieved the second-highest success rate (72.1\%).
Notably, \sys successfully generated a patch for \cc{pdfbox}'s 0-day vulnerability in the AIxCC final.
After the competition, we found that this patch is identical to the one adopted by the project maintainer.
We believe these results demonstrate that \sys can be a practical solution for real-world continuous fuzzing.

Our work makes the following contributions:
\begin{itemize}
    \item \PP{A continuous vulnerability repair (CVR) system integrated with continuous fuzzing.}
    We design \sys as a CVR system tightly integrated with continuous fuzzing.
    Unlike prior studies, \sys is designed to continuously manage the inflow of crash reports, perform deduplication, and generate patches.

    \item \PP{Novel techniques}
    To that end, we introduce several novel techniques, including an ensemble-based approach, two-phase deduplication, and \fp orchestration.
    Our techniques make \sys operationally robust and effective in continuous fuzzing.

    \item \PP{Real-world impact}
    We evaluated \sys on the AIxCC benchmark and in the AIxCC final competition.
    Overall, \sys outperformed the baseline system in the benchmark and in the final competition.
    More interestingly, \sys successfully generated a patch for \cc{pdfbox}'s 0-day vulnerability, demonstrating its practicality.
\end{itemize}

\section{Background}
\label{s:background}
\subsection{Automatic Vulnerability Repair (AVR)}
\label{ss:apr}

Automatic Vulnerability Repair (AVR), also known as Automated Program Repair (APR), is a technique for automatically fixing software vulnerabilities~\cite{appatch,san2patch,pearce2023examining,repairagent,zhang2024fixingsecurityvulnerabilitiesai,genprog,senx,extractfix,codex_old,AlphaRepair}.
For that, AVR systems typically analyze a defect to identify relevant code locations, generate candidate patches through heuristic search or LLMs, and validate the generated candidates.
This validation consists of three steps: 1) compiling the patched program, 2) reproducing the vulnerability, and 3) verifying whether the patched program maintains original functionality. To this end, these systems often use the program's test suite (e.g., unit or regression tests).
If a patch passes this validation, we refer to it as a \textit{plausible patch}. However, plausibility does not guarantee correctness.
This is because, in large and complex software, it is extremely difficult to automatically validate its correctness. Therefore, there could be a plausible, but incorrect patch; it just suppresses the symptom (e.g., removing a critical function) without addressing the underlying root cause. 

\subsection{AI Cyber Challenge (AIxCC)}
\label{ss:aixcc}

AIxCC (AI Cyber Challenge)~\cite{aixcc-website} is a two-year competition hosted by the U.S. Defense Advanced Research Projects Agency (DARPA).
The competition aims to advance fully autonomous techniques for securing large-scale Open-Source Software (OSS) under realistic operational constraints, including finite computational resources and LLM usage budgets.

\PP{Cyber Reasoning System (CRS)}
In AIxCC, participants must develop autonomous Cyber Reasoning Systems (CRSs) that discover vulnerabilities and generate patches without human intervention. AIxCC also challenges the participants to develop their system with high operational robustness.
After participants deploy their systems to the provided cloud infrastructure, they can no longer interact with them.
Thus, any failure in the system can become a single points of failure that stop the system from operating continuously.

A CRS must operate as a continuously running system.
During the competition, the organizers dynamically assign Challenge Projects (CPs) to each CRS.
Then, the CRS must analyze the target codebase, identify vulnerabilities, and generate patches.
For that, AIxCC adopts OSS-Fuzz as its underlying fuzzing framework.
Accordingly, we need to integrate our patching system with this continuous fuzzing pipeline.

\PP{Scoring}
AIxCC evaluates CRSs based on two primary criteria: bug discovery and patch generation~\footnote{AIxCC also includes additional ways to evaluate the CRS (e.g., bundle or SARIF assessment).
However, these are not relevant to our work, so we simply ignore them.}.
To achieve a score for bug discovery, a CRS must submit a Proof-of-Vulnerability (PoV), an input that triggers the vulnerability under the given harness.
For patch generation, the CRS must submit a unified diff file that repairs the vulnerability.
It is worth noting that organizers manually validate submitted patches after the competition ends.
Therefore, the CRS must produce semantically correct patches, not just plausible ones (i.e., patches that suppress the PoV and pass functional tests).

\PP{Challenge Project (CP)}
Each CP consists of a real-world OSS codebase written in C or Java, including widely deployed projects such as \cc{openssl} and \cc{log4j}.
CPs include synthetic vulnerabilities that are manually crafted by domain experts to reflect real-world vulnerabilities~\cite{aixcc-scoring}.
As CPs are based on real-world OSS, they may contain zero-day vulnerabilities.
AIxCC also contributes to the score if the CRS can discover and patch them.

CPs are instantiated in one of two analysis modes: full mode or delta mode.
In full mode, a CRS needs to analyze the entire codebase,
while in delta mode, it needs to analyze the code changes with the given diffs.
In delta mode, a target vulnerability must manifest only after the diffs are applied.
However, this does not imply that the vulnerability resides in the modified code.
For example, a diff may just activate a feature that exposes a latent vulnerability.
Using these two modes, the organizers can evaluate the CRS's capabilities in comprehensive and incremental analysis.

\section{Overview}
\label{s:overview}

In this section, we discuss the design goals of \sys and our approach to achieving them.

\subsection{Design Goals}

\PP{Effectiveness}
The first design goal of \sys is effectiveness. The system must generate patches for given vulnerabilities effectively. In practice, this effectiveness is commonly defined by how well a system produces plausible patches. Although our ultimate objective is to generate correct patches, as mentioned before, it is often infeasible to verify correctness in an automated manner, particularly for large-scale software. Consequently, \sys primarily focuses on generating plausible patches while considering correctness whenever possible.

\PP{Efficiency}
The second design goal is efficiency. This goal is inherently in a trade-off relationship with effectiveness; if we allocate more resources (e.g., servers or LLM tokens), we can improve patch quality. While we prioritize effectiveness, we still need to consider efficiency. For instance, in continuous fuzzing,
fuzzers often generate duplicate crashes from the same underlying root cause. In such cases, it is inefficient to generate patches for redundant crashes. Similarly, it is also not desirable to repeatedly report the same patch to the developer. Thus, \sys needs to handle such redundancy issues efficiently.

\PP{Operational Robustness}
The third design goal is operational robustness. Beyond simple automation, \sys must remain stable and continuously operational over multiple days (e.g., over a week). For that, \sys should operate across diverse projects, build configurations, and vulnerability types.
Moreover, \sys should support new projects without any prior knowledge~\cite{ossfuzz}.
In this context, the system must account for unexpected failures—such as out-of-memory errors or network disruptions—that may arise during operation but are rarely considered in existing AVR systems.

\subsection{Our Approach}

\PP{Ensemble of agents}
To achieve effectiveness and operational robustness, \sys adopts an ensemble-of-agents approach.
We refer to this design as \emph{buckets of agents}, inspired by the buckets of models in ensemble learning.
In this design, \sys uses multiple agents to generate candidate patches to discover a plausible solution.
Through this ensemble, \sys can effectively generate patches by broadening the search space beyond what a single agent can explore.
This also helps for operational robustness; we can isolate the failure of a single agent from the entire system.

To support efficient development of agents, \sys provides a framework called \syscrete.
\syscrete abstracts common functionalities (e.g., code search, environment setup, build automation, and test orchestration) that are shared by all agents;
This allows agent developers to focus on agent-specific logic.
This also ensures a consistent execution environment for diverse agents.

\PP{Two-phase deduplication}
For efficiency, \sys handles two types of duplicates: crash-side duplicates and patch-side duplicates.
Crash-side duplicates occur when multiple crashes originate from the same root cause, 
while patch-side duplicates arise when \sys generates multiple yet incomplete patches for a single root cause.
It is important to handle them efficiently for continuous fuzzing environments.

To address these issues, \sys employs two-phase deduplication.
Our key technique is patch-based deduplication; 
it is based on the observation that if we have a correct patch for a certain root cause, 
we can utilize it to group related crashes and candidate patches.
We describe further details in \autoref{ss:deduplication}.

\PP{\fp orchestration}
\sys also employs an orchestration mechanism called \fp (First-come first-served, Preference-based, and Provider-aware).
In particular, \sys runs multiple workers in parallel, in which each worker runs multiple agents sequentially.
We assigned agents to each worker, considering both LLM providers and preferences.
Moreover, \sys uses a simple yet robust FCFS (First-Come First-Served) policy to submit a patch among multiple workers.
We discuss more details in \autoref{ss:hybrid_scheduling}.

\section{Design}
\label{s:design}

\begin{figure*}[ht]
    \centering
    \includegraphics[width=0.9\linewidth]{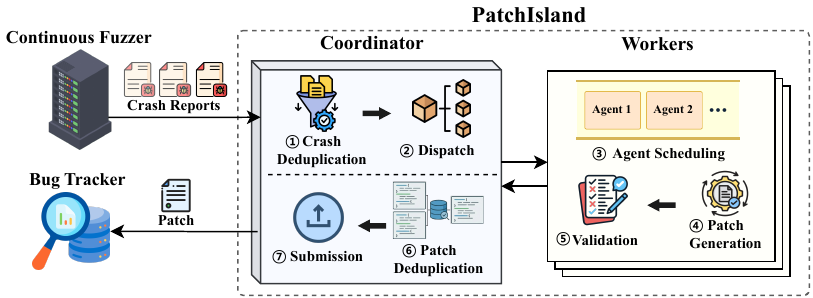}
    \caption{
        Workflow of patch generation. \sys operates between the continuous fuzzing pipeline and the bug tracker, receiving crash reports and delivering generated patches to developers. The \coordinator delegates patch generation to distributed \workers.
    }
    \label{fig:architecture}
\end{figure*}

In this section, we discuss the design of \sys.

\subsection{Overview}
\label{ss:operational_arch}

\PP{Distributed Architecture}
For operational robustness and scalability,
\sys adopts a distributed architecture operating in a cloud environment (i.e., \cc{k8s}).
\sys follows a \coordinator-\worker model; it has a single coordinator and multiple workers.
In particular, \coordinator manages system-level state and dispatches tasks, 
while \workers handle resource-intensive patch generation.
This design allows \sys to remain stable even when \worker fails and to scale horizontally by adding more \workers.

\PP{Workflow}
As shown in \autoref{fig:architecture}, \sys operates as follows: \WC{1} Upon
receiving a crash, \coordinator deduplicates crash reports to filter out
redundant ones (Phase~1 of \autoref{ss:deduplication}). If the crash is unique, \WC{2} it
dispatches this task to an available \worker. \WC{3} Each \worker runs
agents sequentially
according to the agent configurations.
\WC{4} Then, each agent generates a candidate patch; if an agent fails to produce a patch, the next
agent starts (detailed in \autoref{ss:hybrid_scheduling}). \WC{5} After an agent
successfully generates a candidate patch, \worker validates it and sends the
result back to \coordinator. \WC{6} \coordinator then deduplicates the
incoming patches to eliminate duplicates (Phase~2 of \autoref{ss:deduplication}). \WC{7} For final review, the
\coordinator submits the validated patch or updates the previously one
if the new patch offers a better fix for the underlying root cause.

\subsection{Two-Phase Deduplication}
\label{ss:deduplication}

\begin{algorithm}[t]
\footnotesize
\DontPrintSemicolon
\SetKwSty{algokeywordsty}
\SetFuncSty{algofuncsty}
\SetArgSty{algoargsty}

\caption{Two-phase deduplication}
\label{alg:patch-dedup}

\SetKwFunction{OnCrashReceived}{OnCrashReceived}
\SetKwFunction{OnPatchGenerated}{OnPatchGenerated}
\SetKwFunction{ResolvedByExistingPatches}{ResolvedByExistingPatches}
\SetKwFunction{AddToDispatchQueue}{AddToDispatchQueue}
\SetKwFunction{StorePatch}{StorePatch}
\SetKwFunction{Rebuild}{Rebuild}
\SetKwFunction{ResolvedByPatch}{ResolvedByPatch}
\SetKwFunction{RemoveFromDispatchQueue}{RemoveFromDispatchQueue}
\SetKwFunction{MergePatch}{MergePatch}

\SetKwProg{Proc}{Procedure}{}{}

\Proc{\OnCrashReceived{crash}}{
    \If{\ResolvedByExistingPatches{crash}}{
        \Return
    }
    \AddToDispatchQueue{crash}\;
}
\vspace{0.8em}

\Proc{\OnPatchGenerated{patch}}{


    \ForEach{crash \textbf{ in } dispatch\_queue}{
        \If{\Not \ResolvedByPatch{patch, crash}}{
            \RemoveFromDispatchQueue{crash}\;
        }
    }

    \ForEach{p \textbf{ in } stored\_patches}{
        \ForEach{crash \textbf{ in } PoVs\_of(p)}{
            \If{\Not \ResolvedByPatch{patch, crash}}{
                \MergePatch{patch, p}\;
            }
        }
    }

    \StorePatch{patch}\;
}
\end{algorithm}

Crash deduplication is one of the primary challenges in designing a CVR system.
While existing fuzzers typically perform their own deduplication~\cite{afl,honggfuzz}, 
they often fail to deduplicate crashes originating from the same root cause.
There have been many research works on crash deduplication~\cite{igor,GPTrace,RETracer,rept,aurora};
however, many of them can be inaccurate as they rely on stack-based heuristics~\cite{clusterfuzz,afl,honggfuzz}.
Others~\cite{igor} are specifically designed for certain fuzzing engines (e.g., AFL++~\cite{afl}),
and some~\cite{rept} assume hardware support (e.g., \cc{Intel PT}~\cite{intel_pt}), which limits their applicability.
As a result, these methods are not suitable for \sys, which needs to robustly operate alongside continuous fuzzing.

To address this limitation, \sys adopts a two-phase deduplication strategy.
Its underlying idea is to use \emph{patch-based deduplication}~\cite{aixcc-scoring}.
This approach assumes that a correct patch, which properly addresses a root cause, 
will also resolve other duplicate crashes, allowing us to group them together.
Moreover, it is efficient to re-run PoVs with the patched binary, making it a practical approach.
As shown in \autoref{alg:patch-dedup}, this deduplication process operates in two phases.

\PP{Phase 1: Crash-side deduplication.} 
\sys deduplicates crashes at two stages during processing.
First, when a crash arrives, \sys checks whether it is a duplicate by re-running it against previously generated patches (Lines 2–3).
If the crash is resolved by one of these patches, \sys classifies it as a duplicate and stops further processing.
Second, after a new patch is generated, \sys re-checks the crashes in the dispatch queue (Lines 7–8).
To do this, \sys re-validates the crashes in the dispatch queue against the new patch.
This step avoids potential race conditions, as \sys may generate a patch for duplicates after a crash has already arrived.

\PP{Phase 2: Patch-side deduplication.} 
\sys also handles duplicates for patches.
To be effective, \sys needs to avoid submitting multiple patches for the same issue. 
Unfortunately, as mentioned before,
we can only automatically determine a patch's plausibility, not its correctness.
As a result, \sys can generate multiple plausible patches but incorrect ones.
To avoid this, \sys verifies whether a new patch can subsume an existing patch whenever it is generated (Line 10--11).
If yes, \sys merges the new patch with the existing one and reports this to the developer.
This is reasonable as the new patch can address more crashes than the previous one (e.g., the current \cc{crash}).
However, this approach also has a risk that it creates a superman patch --- a complex patch that addresses multiple issues at once.
We will discuss this further in the discussion section (see \autoref{s:discussion}).

\subsection{\fp Orchestration}
\label{ss:hybrid_scheduling}
\begin{figure}[t]
    \centering
    \includegraphics[width=\linewidth]{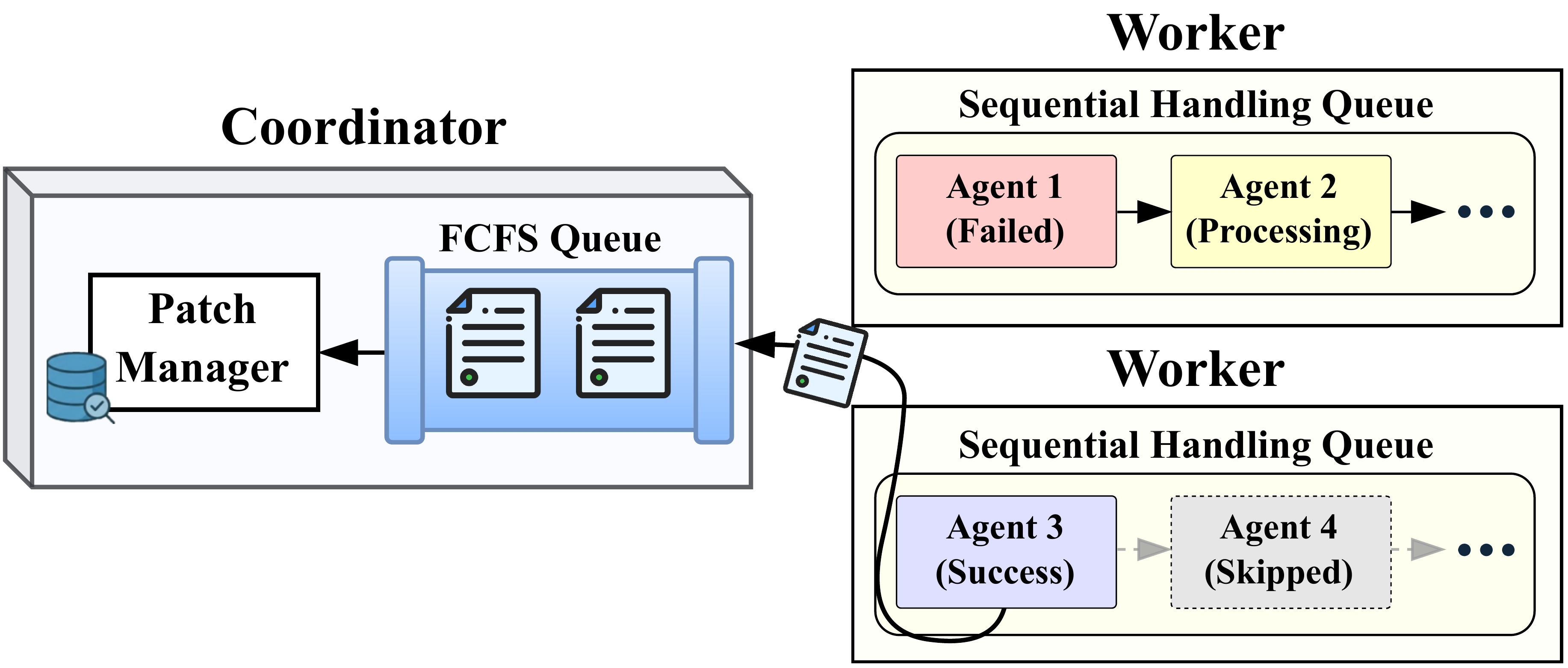}
    \caption{\emph{FP$^2$} orchestration strategy with Ensemble}
    \label{fig:scheduling}
\end{figure}

\sys introduces \fp (FCFS, Preference-based, and Provider-aware) orchestration to schedule agents.
In particular, \sys runs multiple agents in parallel, in which each agent runs sequentially.
For that, we determine their order considering LLM providers and preferences.
Our design is aligned with our design goals in \autoref{s:overview}.
Particularly, we found that many existing approaches~\cite{dei, papr,mahmud2025enhancingllmcodegeneration,aurora} 
do not consider operational robustness, making them unsuitable for \sys. 
In the following, we discuss more details.

\PP{Result Selection Policy: FCFS}
\sys adopts a simple First-Come, First-Served (FCFS) policy for selecting
candidate patches. In contrast to prior works, which use more sophisticated
mechanisms (e.g., LLM-as-a-judge~\cite{dei}),
FCFS avoids additional coordination and reduces the impact of \worker failures.
As a result, this simple policy is more favorable in continuous fuzzing.

\PP{Sequential Execution: Preference-based Allocation}
In sequential execution, we prioritize agents based on our preferences, which are determined by the empirical likelihood of producing correct patches.
During development, we observed that some agents are more likely to produce correct patches, whereas others generate patches quickly but are not truly correct (e.g., function removal or early returns).
Such patches are often considered plausible when a project has limited functional tests.
In such cases, if we execute all agents in parallel using the FCFS policy, fast but low-quality agents tend to dominate the results.
To mitigate this, \sys prioritizes agents that are more likely to produce correct patches.
For simplicity and operational robustness, 
we use a statically defined preference for this purpose.

\PP{Parallel Execution: Provider-aware Allocation}
In parallel execution, we consider LLM providers to avoid provider-level throttling.
Because each provider enforces usage quotas (e.g., Token Per Minute or API rate
limits), if agents share the same provider, they may be throttled simultaneously.
To avoid this, \sys schedules agents from the different providers in parallel,
thereby reducing the impact of provider-level throttling.

\subsection{\syscrete: A framework for agent development}
\label{ss:syscrete_implementation}

We provide a framework, \syscrete, to facilitate agent development.
In particular, \syscrete provides a variety of shared tools, including code search, environment setup, build automation, and test orchestration. By using \syscrete, developers can focus on their unique workflow and core logic without redundant effort, enabling rapid development.

\subsubsection{Core modules}
\syscrete supports the following core modules:

\PP{OSS-Fuzz Environment} This module provides managed interfaces for interacting with projects. In particular, it provides a way to build, test, apply patches, and reproduce PoVs.

\PP{Evaluator} This module assesses the quality of patches. For that, it verifies compilation, reproduction, and functional correctness to ensure that patches mitigate vulnerabilities without breaking the original functionality.

\PP{Fault localizer} This module identifies the fault location from crash logs.
It offers several approaches for this purpose, including simple stack-based analysis and more sophisticated analysis techniques inspired by CodeRover-S~\cite{autocoderover}.

\PP{Code retriever} This module supports code navigation, enabling agents to query code elements such as functions or variables. It leverages tools such as Tree-Sitter~\cite{treesitter} and ctags~\cite{ctags}.

\PP{Analyzers} This module provides reusable analyzers for crash logs, stack traces, and commits, allowing agents to extract information.

\subsubsection{Advanced features} \syscrete also provides: 

\PP{Environment pool}
An environment pool is a collection of isolated build and execution environments that \syscrete manages and reuses across repair attempts to reduce build overhead.
In continuous repair, repeated builds for testing and validation often become a bottleneck.
To address this issue, \syscrete reuses prebuilt artifacts, such as compiled binaries and libraries, thereby avoiding redundant builds.
For operational robustness, \syscrete maintains heterogeneous environments within the pool.
For example, some environments enable \cc{ccache} to accelerate builds, while others disable it to handle cases where \cc{ccache} is unreliable.

\PP{Cache everywhere}
To improve efficiency, \syscrete adopts a cache-everywhere strategy. It attempts to cache all computationally expensive yet deterministic results. These include crash reproduction, patch validation, and static analyses. This mechanism helps developers to focus on the core logic of their agents, without worrying about redundant computations. For example, a developer can validate one patch multiple times without considering the overhead, as \syscrete will use cached results if available.

\PP{Special handling for certain bug types}
\syscrete also includes specialized handling for certain types of bugs. For instance, stack overflow crashes in Java often produce repetitive stack frames, which hinder LLM-based analysis. To mitigate this, \syscrete preprocesses stack traces to remove redundant frames. Similarly, timeout bugs in Java typically do not generate stack traces, complicating the initial analysis. To address this, \syscrete leverages \cc{jstack}~\cite{jstack} to capture stack traces when a timeout occurs. In particular, if the crash is a Java timeout, \syscrete reproduces it while periodically capturing 30 stack traces with \cc{jstack}. These stack traces can reveal the loop location, which can help agents to patch the bug.

\subsubsection{Example: Integrating existing agents}
Thanks to \syscrete, we can significantly reduce the engineering effort needed to implement agents.
This also applies to integrating existing agents into \sys.
As shown in \autoref{tab:agent_integration_effort}, we can integrate diverse agents with only a few hundred lines of code, including prompts.

\begin{table}[t]
    \footnotesize
    \centering
    \caption{The number of new lines of code required to integrate various agents into \sys.}
    \label{tab:agent_integration_effort}
    \begin{adjustbox}{max width=\linewidth}
        \begin{tabular}{llc}
            \toprule
            \textbf{Agent} & \textbf{Type} & \textbf{LoC} \\
            \midrule
            Codex          & Commercial      & 143 \\
            Claude Code    & Commercial      & 390 \\
            \syssweagent   & Open Source     & 570 \\
            \sysaider      & Open Source     & 646 \\
            \bottomrule
        \end{tabular}
    \end{adjustbox}
\end{table}

\subsection{In-house Agents}

\label{s:agents}

In this section, we describe five in-house agents used in \sys at a high level.
Our goal is to increase the diversity of our agents.
Thus, we employ different methodologies and models for each agent.
Due to space constraints, we focus on their roles and design characteristics,
and refer to our technical report~\cite{techreport} for detailed implementations.

\PP{MultiRetrieval}
An agent that follows a two-step workflow: it first analyzes the crash to gather relevant code context, then iteratively generates and refines patches.
It relies on multiple code-query mechanisms to expand context as needed during patch synthesis, enabling steady improvement across iterations.

\PP{Vincent}
A three-stage workflow consisting of root-cause analysis, project-specific property analysis, and patch generation.
By incorporating properties inferred from the codebase---such as expected behaviors or constraints---Vincent produces patches that address semantic requirements rather than superficial symptom fixes.

\PP{Martian}
A two-part system where separate agents handle crash analysis and patch generation.
This division allows each stage to specialize: the analysis component focuses solely on locating the fault, while the patching component applies targeted edits using a tailored toolset, such as a language server (\eg, \cc{clangd}~\cite{clangd} and \cc{Eclipse JDT LS}~\cite{eclipsejdtls}) to improve code navigation.

\PP{ClaudeLike}
A patching framework in which the patching agent repeatedly calls an on-demand analysis agent whenever more context or clarification is required.
This interactive pattern enables fine-grained information requests during diff construction, improving the accuracy and completeness of complex patches.

\PP{Prism}
A cyclic workflow that rotates between analysis, patching, and evaluation agents.
Each iteration refines the system's understanding of the bug and the patch's correctness, allowing the process to gradually converge on a validated solution despite context-size or complexity challenges.

\section{Implementation}
\label{s:implementation}

To implement \sys, we developed approximately 40,000 lines of Python code.
In the following, we describe the implementation details of \sys, which is interesting to discuss.

\PP{Artifacts Sharing}
\sys employs a pre-initialized environment to reduce redundant work across agents.
When \sys receives project information, it initializes the target OSS-Fuzz project once and prepares all required build artifacts, including the Docker image, containers, and populated build cache.
All agents inherit this environment, avoiding repeated initialization steps.

\PP{Build Caching for C}
For C projects, \sys leverages existing build caching tools---primarily \cc{ccache}~\cite{ccache}.
In a preliminary evaluation on 110 buildable OSS-Fuzz C projects, \cc{ccache} applied to 107 of them (approximately 97.3\%).
Across these projects, we observed a 55\% time-weighted average reduction in build time.
These optimizations come directly from the underlying toolchains and are orthogonal to the design of \sys.

\PP{Dependency Sharing for Java}
For Java projects, \sys reduces build overhead by sharing the local Maven dependency repository across agents.
In our evaluation, projects that received this optimization achieved a 66\% reduction in build time by eliminating repeated dependency downloads.

\PP{Kubernetes Integration}
We deploy the \coordinator and four \workers as separate Kubernetes pods. This
setup isolates failures---if a pod crashes, it restarts independently without
affecting the others. All components communicate through predefined APIs over
the network and share a project repository via a mounted volume.

\section{Evaluation}
\label{s:eval}
In this section, we attempt to answer the following questions:

\begin{itemize}
    \item How is effective \sys compared to other systems in the AIxCC competition? (\autoref{ss:rq_comparison})
    \item How effective is \sys's ensemble strategy compared to a single-agent strategy? (\autoref{ss:compare_single_vs_ensemble})
    \item How effective is \sys's \fp orchestration? (\autoref{ss:rq_scheduling})
\end{itemize}

\PP{Experimental Setup}
We ran all experiments on a machine equipped with an Intel Xeon Gold 6346 CPU (16 cores) at 3.10 GHz, 128 GB RAM, and 1 TB of storage, running Ubuntu 22.04 LTS.
As mentioned earlier, \sys utilizes diverse LLMs, including \cc{GPT o4-mini}, \cc{Claude Sonnet 4}, and \cc{Gemini 2.5 Pro}.

\PP{Dataset}
\begin{table}[t]
    \centering
    \caption{%
        Dataset overview of the CP vulnerabilities released after the conclusion of the AIxCC competition.
        This table summarizes 92 vulnerabilities selected from the publicly released CP dataset,
        restricted to cases that include an PoV).
        Vulnerabilities are aggregated per project and dataset type (delta or full).
        The complete list of vulnerabilities and their detailed descriptions are provided in
        \autoref{tab:appendix-cp-vulns}.
    }
    \label{tab:dataset-summary}
    \setlength{\tabcolsep}{4pt}
    \renewcommand{\arraystretch}{0.9}
    \small
    \begin{adjustbox}{max width=\linewidth, max height=6.5cm}
        \begin{tabular}{lcrrr}
            \toprule
            \textbf{Project Name}               & \textbf{Language} & \textbf{Delta} & \textbf{Full} & \textbf{Total} \\
            \midrule
            Apache Commons Compress             & Java              & 7              & 4             & 11             \\
            Apache POI                          & Java              & 2              & 5             & 7              \\
            curl                                & C                 & 7              & 1             & 8              \\
            Dicoogle                            & Java              & 0              & 0             & 0              \\
            dav1d                               & C                 & 0              & 1             & 1              \\
            dcm4che                             & Java              & 0              & 0             & 0              \\
            FreeRDP                             & C                 & 3              & 2             & 5              \\
            healthcare-data-harmonization (hdh) & Java              & 0              & 0             & 0              \\
            Apache HertzBeat                    & Java              & 0              & 0             & 0              \\
            jsoup                               & Java              & 0              & 0             & 0              \\
            Little CMS (lcms)                   & C                 & 0              & 2             & 2              \\
            libavif                             & C                 & 1              & 0             & 1              \\
            libexif                             & C                 & 5              & 0             & 5              \\
            libxml2                             & C                 & 4              & 2             & 6              \\
            Apache Log4j 2                      & Java              & 1              & 0             & 1              \\
            Mongoose                            & Java              & 2              & 1             & 3              \\
            nDPI                                & C                 & 0              & 0             & 0              \\
            OpenSSL                             & C                 & 0              & 0             & 0              \\
            PDFBox                              & Java              & 1              & 8             & 9              \\
            shadowsocks-libev                   & C                 & 0              & 5             & 5              \\
            systemd                             & C                 & 0              & 4             & 4              \\
            Tika                                & Java              & 6              & 5             & 11             \\
            Wireshark                           & C                 & 6              & 6             & 12             \\
            XZ                                  & C                 & 0              & 1             & 1              \\
            \midrule
            \textbf{Total}                      &                   & \textbf{45}    & \textbf{47}   & \textbf{92}    \\
            \bottomrule
        \end{tabular}
    \end{adjustbox}
    \normalsize
\end{table}

We evaluate \sys using the AIxCC challenges for the final round.
This dataset is built on 24 real-world open-source software (e.g., Apache Commons Compress, curl, etc.) with 92 synthetic vulnerabilities.

We believe this dataset is well-suited for evaluating an AVR system for the following reasons:
First, this dataset is manually crafted by domain experts to reflect real-world vulnerabilities.
For that, many of the vulnerabilities are inspired by previously discovered real-world vulnerabilities (e.g., CVE-2016-1898 and CVE-2023-42503) or represent issues that could arise in practice (e.g., CWE-193: Off-by-one Error~\cite{CWE-193}).
Second, this dataset spans two programming languages: C and Java. This allows us to assess the system's performance across different programming languages.
Third, this dataset contains diverse vulnerability types. In more detail, it includes not only memory corruptions (\eg, buffer overflows, use-after-free, etc.), but also various logic bugs (\eg, path traversal, zip slip, and OS command injection).
Finally, this dataset also provides important evaluation artifacts.
In particular, it includes PoVs, functional tests, and ground-truth patches, allowing us to evaluate the system's end-to-end performance.

\subsection{Comparison to Other Systems}
\label{ss:rq_comparison}
\begin{table}[t]
    \centering
    \caption{
        Comparison of \sys with other AIxCC competition systems on 92 PoVs.
        This table summarizes final patching performance against publicly available baseline results from the competition.
    }
    \label{tab:patch_performance_official_per_prj-comparison}
    \newcommand{\RowHead}[1]{\vspace{0.1pt}\textbf{\shortstack{#1}}}
\newcommand{\ColHead}[1]{\textbf{\shortstack{#1}}}
\newcommand{\AGENT}[1]{\rule{0pt}{1.5em}#1\rule[-1.3em]{0pt}{0pt}}
\newcommand{\RATIO}[2]{(\ratio{#1}{#2})}
\newcommand{\PAD}[0]{\hspace{14pt}}
\rowcolors{2}{gray!20}{white}
\setlength{\heavyrulewidth}{1pt}
\setlength{\lightrulewidth}{1pt}   
\setlength{\abovetopsep}{1pt}
\setlength{\belowrulesep}{0pt}
\setlength{\aboverulesep}{0pt}
\setlength{\arrayrulewidth}{1pt}
\setlength{\tabcolsep}{4pt}
\begin{adjustbox}{max width=\linewidth, max height=\linewidth}
    \begin{tabular}{cclclcl}
        \toprule
        \RowHead{Project Name (\cc{Language})} & \multicolumn{2}{c}{\PAD\RowHead{\baselinea}} & \multicolumn{2}{c}{\PAD\RowHead{\baselineb}} & \multicolumn{2}{c}{\PAD\RowHead{\sys}}                                                                                \\
        \midrule
        \ColHead{commons-compress (\cc{java})} & \PAD  5 / 11                                 & \RATIO{5}{11}                           & \PAD 7 / 11                           & \RATIO{7}{11}            & \PAD 10 / 11            & \RATIO{10}{11}            \\
        \ColHead{poi (\cc{java})}              & \PAD  1 / 7                                  & \RATIO{1}{7}                            & \PAD 4 / 7                             & \RATIO{4}{7}             & \PAD 5 / 7             & \RATIO{5}{7}             \\
        \ColHead{curl (\cc{c})}                & \PAD  0 / 8                                  & \RATIO{0}{8}                            & \PAD 3 / 8                             & \RATIO{3}{8}             & \PAD 7 / 8             & \RATIO{7}{8}             \\
        \ColHead{dav1d (\cc{c})}               & \PAD  0 / 1                                  & \RATIO{0}{1}                            & \PAD 0 / 1                             & \RATIO{0}{1}             & \PAD 0 / 1             & \RATIO{0}{1}             \\
        \ColHead{freerdp (\cc{c})}             & \PAD  1 / 5                                  & \RATIO{1}{5}                            & \PAD 2 / 5                             & \RATIO{2}{5}             & \PAD 5 / 5             & \RATIO{5}{5}             \\
        \ColHead{lcms (\cc{c})}                & \PAD  0 / 2                                  & \RATIO{0}{2}                            & \PAD 2 / 2                             & \RATIO{2}{2}             & \PAD 2 / 2             & \RATIO{2}{2}             \\
        \ColHead{libavif (\cc{c})}             & \PAD  0 / 1                                  & \RATIO{0}{1}                            & \PAD 1 / 1                             & \RATIO{1}{1}             & \PAD 1 / 1             & \RATIO{1}{1}             \\
        \ColHead{libexif (\cc{c})}             & \PAD  1 / 5                                  & \RATIO{1}{5}                            & \PAD 2 / 5                             & \RATIO{2}{5}             & \PAD 5 / 5             & \RATIO{5}{5}             \\
        \ColHead{libxml2 (\cc{c})}             & \PAD  2 / 6                                  & \RATIO{2}{6}                            & \PAD 5 / 6                             & \RATIO{5}{6}             & \PAD 6 / 6             & \RATIO{6}{6}             \\
        \ColHead{log4j2 (\cc{java})}           & \PAD  1 / 1                                  & \RATIO{1}{1}                            & \PAD 0 / 1                             & \RATIO{0}{1}             & \PAD 1 / 1             & \RATIO{1}{1}             \\
        \ColHead{mongoose (\cc{c})}            & \PAD  0 / 3                                  & \RATIO{0}{3}                            & \PAD 0 / 3                             & \RATIO{0}{3}             & \PAD 2 / 3             & \RATIO{2}{3}             \\
        \ColHead{pdfbox (\cc{java})}           & \PAD  1 / 9                                  & \RATIO{1}{9}                            & \PAD 6 / 9                             & \RATIO{6}{9}             & \PAD 8 / 9             & \RATIO{8}{9}             \\
        \ColHead{shadowsocks (\cc{c})}         & \PAD  0 / 5                                  & \RATIO{0}{5}                            & \PAD 4 / 5                             & \RATIO{4}{5}             & \PAD 5 / 5             & \RATIO{5}{5}             \\
        \ColHead{systemd (\cc{c})}             & \PAD  0 / 4                                  & \RATIO{0}{4}                            & \PAD 3 / 4                             & \RATIO{3}{4}             & \PAD 4 / 4             & \RATIO{4}{4}             \\
        \ColHead{tika (\cc{java})}             & \PAD  3 / 11                                 & \RATIO{3}{11}                           & \PAD 7 / 11                           & \RATIO{7}{11}            & \PAD 10 / 11            & \RATIO{10}{11}            \\
        \ColHead{wireshark (\cc{c})}           & \PAD  1 / 12                                 & \RATIO{1}{12}                           & \PAD 10 / 12                           & \RATIO{10}{12}           & \PAD 12 / 12           & \RATIO{12}{12}           \\
        \ColHead{xz (\cc{c})}                  & \PAD  0 / 1                                  & \RATIO{0}{1}                            & \PAD 1 / 1                             & \RATIO{1}{1}             & \PAD 1 / 1             & \RATIO{1}{1}             \\
        \midrule
        \RowHead{Totals}                       & \PAD \RowHead{16 / 92}                       & \RowHead{\RATIO{16}{92}}                & \PAD \RowHead{57 / 92}                 & \RowHead{\RATIO{57}{92}} & \PAD \RowHead{84 / 92} & \RowHead{\RATIO{84}{92}} \\
        \bottomrule
    \end{tabular}
\end{adjustbox}

\end{table}

\begin{figure}[t]
    \centering
    \includegraphics[width=\linewidth]{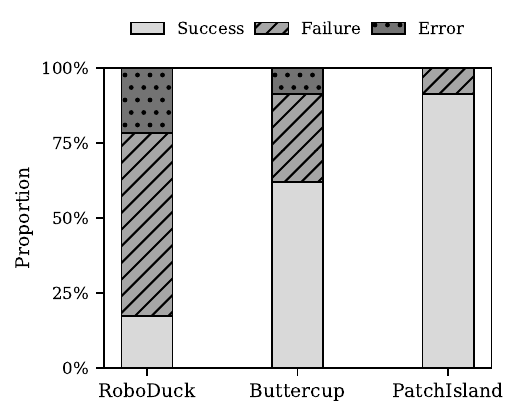}
    \caption{
        Comparison of patching outcomes between \sys and other AIxCC systems, showing the number of successfully patched vulnerabilities and the proportion of error cases for each system.
    }
    \label{fig:patch_success_comparison}
\end{figure}

To demonstrate \sys's effectiveness, we compare it against other AIxCC competition systems.

\PP{Baseline}
For baseline comparison, we selected \baselinea and \baselineb, the top-performing systems among all competition submissions in terms of patching effectiveness.
For baseline systems, we use the default configurations they used, assuming their AIxCC submissions reflect their best-performing settings.
We used PoVs from the official AIxCC dataset as inputs for all systems.

\PP{Threats to Validity}
In this paragraph, we discuss potential threats to validity in our evaluation. 
The baseline systems were originally designed to run the full AIxCC workflow (e.g., fuzzing, patching, and reporting). 
For our evaluation, we isolated and evaluated only their patch generation components. 
Although we made careful efforts to preserve their original behaviors, 
this isolation may have altered implicit dependencies or assumptions in their full workflow. 
To mitigate this threat, we also compare \sys with these systems in an end-to-end setting, as described in \autoref{s:post_mortem}.

\PP{Results}
\autoref{tab:patch_performance_official_per_prj-comparison} compares the number of successfully patched vulnerabilities by \sys and the two baseline systems. 
Overall, \sys outperforms both baseline systems; 
\sys successfully patches 84 out of 92 vulnerabilities, while \baselinea and \baselineb patch 57 and 16 vulnerabilities, respectively.

\autoref{fig:patch_success_comparison} further visualizes these results by illustrating the proportion of successes, failures, and errors for each system.
We define success, failure, and error as follows:
\begin{itemize}
  \item \textbf{Success} refers to a \textit{plausible patch} that prevents PoV reproduction and passes the functional tests.
  \item \textbf{Failure} refers to cases where the system terminates without producing a patch due to internal limits (e.g., the maximum number of trials or timeout)
  \item \textbf{Error} refers to system-level failures (e.g., build failures or PoV reproduction failures) that block further system execution as intended.
\end{itemize}
It is worth mentioning that \sys exhibits \emph{zero} errors, whereas \baselinea and \baselineb show 21.7\% and 8.7\% errors, respectively.
This is because of \sys's ensemble-based approach.
Even though \sys's individual agents may fail, other agents can continue operating, making the overall system robust.
For instance, \sysmartian can fail due to LSP (Language Server Protocol) failure; however, other agents that do not rely on the LSP can still proceed.

\subsection{Effectiveness of the Ensemble Strategy}
\label{ss:compare_single_vs_ensemble}

To evaluate our ensemble approach, we compare two strategies: (1) repeatedly executing a single agent and (2) using an ensemble of diverse agents.
This comparison examines whether ensemble diversity offers advantages over simple repetition.

\PP{Preliminary Study}
Before the main experiment, we conducted a preliminary study to determine the number of runs ($k$), required for each agent’s results to converge. 
In the main experiment, we use this number to evaluate our approach, mitigating the non-determinism of LLMs.
Specifically, we repeatedly executed each agent and measured how the outcomes changed as the number of runs increased. For that, we cumulatively tracked the number of successful patches across runs.

\begin{figure}[t]
    \centering
    \includegraphics[width=\linewidth]{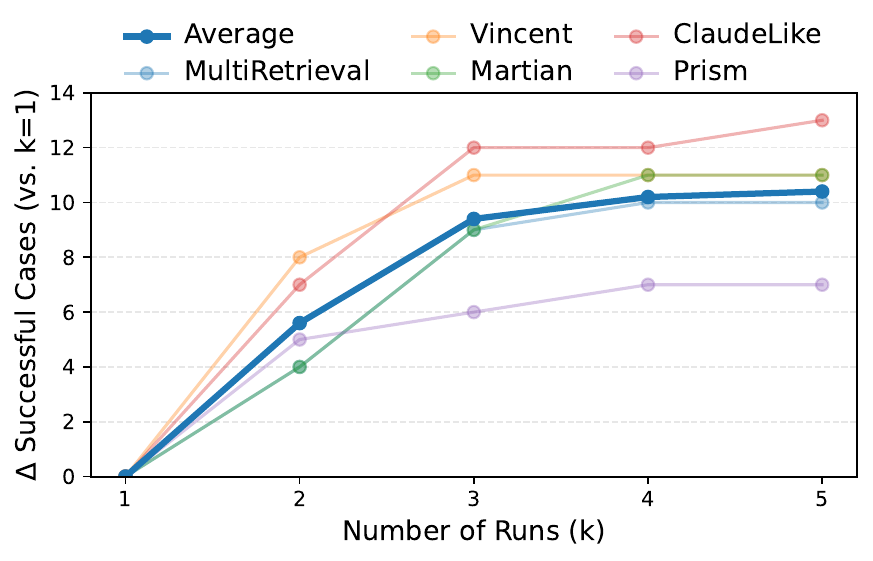}
    \caption{
    Effect of repeated executions for a non-deterministic agent.
    The performance gains rapidly saturate, indicating diminishing returns from additional executions.
    }

    \label{fig:saturated_result}
\end{figure}

\autoref{fig:saturated_result} shows the results of this preliminary study.
In summary, we found that most agents converge after \emph{five} executions.
While all agents return the same results at the fifth execution, the only exception is \sysclaudelike, which discovers one additional successful patch at the fifth execution.
This implies that agents do not fully converge even after five executions;
however, we chose five as the number of runs ($k$ = 5) by considering both cross-agent trends and LLM costs.

\PP{Results} 
Our ensemble-based approach mitigates the limitations of individual agents by combining their own strengths.
As shown in \autoref{tab:patch_performance_official_per_prj}, our ensemble approach --- reported as the best-of-$N$ --- could repair up to 87 vulnerabilities by leveraging agent diversity.
\sysprism and \sysmr are the highest-performing individual agents, while \sysclaudelike and \sysvincent show lower overall performance.
Nevertheless, \sysvincent and \sysclaudelike fixed two and one vulnerabilities that \sysprism and \sysmr failed to repair, respectively.

\begin{table*}[t]
    \centering
    \caption{
        Results of five-run evaluations for five agents on 92 AIxCC PoVs.
        Each cell indicates whether an agent produced a successful patch in at least one of the five identical runs.
        The rightmost column represents an upper-bound reference showing whether any agent succeeded for each PoV.
    }
    \label{tab:patch_performance_official_per_prj}
        \newcommand{\RowHead}[1]{\textbf{\shortstack{#1}}}
    \newcommand{\ColHead}[1]{\textbf{\shortstack{#1}}}
    \newcommand{\AGENT}[1]{\rule{0pt}{1.5em}#1\rule[-1.3em]{0pt}{0pt}}
    \newcommand{\RATIO}[2]{(\ratio{#1}{#2})}
    \newcommand{\PAD}[0]{\hspace{14pt}}
    \rowcolors{4}{gray!20}{white}
    \setlength{\heavyrulewidth}{1pt}
    \setlength{\lightrulewidth}{1pt}   
    \setlength{\abovetopsep}{1pt}
    \setlength{\belowrulesep}{0pt}
    \setlength{\aboverulesep}{0pt}
    \setlength{\arrayrulewidth}{1pt}
    \setlength{\tabcolsep}{4pt}
    \begin{adjustbox}{max width=\linewidth, max height=\linewidth}
        \begin{tabular}{cclclclclcl|cl}
            \toprule
            \multirow{2}{*}{\RowHead{Project Name (\cc{Language}) }} & \multicolumn{10}{c|}{\RowHead{Agents}}           & \multicolumn{2}{c}{\multirow{2}{*}{\RowHead{best-of-N}}}                                                                                                                                                                                                                                                                                                                             \\
            \cmidrule(lr){2-11}                                      & \multicolumn{2}{c}{\PAD\RowHead{\sysclaudelike}} & \multicolumn{2}{c}{\PAD\RowHead{\sysmartian}}                 & \multicolumn{2}{c}{\PAD\RowHead{\sysmr}} & \multicolumn{2}{c}{\PAD\RowHead{\sysprism}} & \multicolumn{2}{c|}{\PAD\RowHead{\sysvincent}} &                                                                                                                                                                                 \\
            \midrule
            \ColHead{commons-compress (\cc{java})}                   & \PAD 9 / 11                                      & \RATIO{9}{11}                                                 & \PAD 10 / 11                             & \RATIO{10}{11}                              & \PAD 10 / 11                                   & \RATIO{10}{11}           & \PAD 10 / 11           & \RATIO{10}{11}           & \PAD 9 / 11            & \RATIO{9}{11}            & 10 / 11           & \RATIO{10}{11}           \\
            \ColHead{poi (\cc{java})}                                & \PAD 3 / 7                                       & \RATIO{3}{7}                                                  & \PAD 4 / 7                               & \RATIO{4}{7}                                & \PAD 5 / 7                                     & \RATIO{5}{7}             & \PAD 4 / 7             & \RATIO{4}{7}             & \PAD 4 / 7             & \RATIO{4}{7}             & 5 / 7             & \RATIO{5}{7}             \\
            \ColHead{curl (\cc{c})}                                  & \PAD 6 / 8                                       & \RATIO{6}{8}                                                  & \PAD 7 / 8                               & \RATIO{7}{8}                                & \PAD 7 / 8                                     & \RATIO{7}{8}             & \PAD 7 / 8             & \RATIO{7}{8}             & \PAD 8 / 8             & \RATIO{8}{8}             & 8 / 8             & \RATIO{8}{8}             \\
            \ColHead{dav1d (\cc{c})}                                 & \PAD 0 / 1                                       & \RATIO{0}{1}                                                  & \PAD 0 / 1                               & \RATIO{0}{1}                                & \PAD 0 / 1                                     & \RATIO{0}{1}             & \PAD 0 / 1             & \RATIO{0}{1}             & \PAD 0 / 1             & \RATIO{0}{1}             & 0 / 1             & \RATIO{0}{1}             \\
            \ColHead{freerdp (\cc{c})}                               & \PAD 5 / 5                                       & \RATIO{5}{5}                                                  & \PAD 4 / 5                               & \RATIO{4}{5}                                & \PAD 5 / 5                                     & \RATIO{5}{5}             & \PAD 5 / 5             & \RATIO{5}{5}             & \PAD 5 / 5             & \RATIO{5}{5}             & 5 / 5             & \RATIO{5}{5}             \\
            \ColHead{lcms (\cc{c})}                                  & \PAD 1 / 2                                       & \RATIO{1}{2}                                                  & \PAD 2 / 2                               & \RATIO{2}{2}                                & \PAD 2 / 2                                     & \RATIO{2}{2}             & \PAD 2 / 2             & \RATIO{2}{2}             & \PAD 2 / 2             & \RATIO{2}{2}             & 2 / 2             & \RATIO{2}{2}             \\
            \ColHead{libavif (\cc{c})}                               & \PAD 1 / 1                                       & \RATIO{1}{1}                                                  & \PAD 1 / 1                               & \RATIO{1}{1}                                & \PAD 1 / 1                                     & \RATIO{1}{1}             & \PAD 1 / 1             & \RATIO{1}{1}             & \PAD 1 / 1             & \RATIO{1}{1}             & 1 / 1             & \RATIO{1}{1}             \\
            \ColHead{libexif (\cc{c})}                               & \PAD 4 / 5                                       & \RATIO{4}{5}                                                  & \PAD 5 / 5                               & \RATIO{5}{5}                                & \PAD 5 / 5                                     & \RATIO{5}{5}             & \PAD 5 / 5             & \RATIO{5}{5}             & \PAD 5 / 5             & \RATIO{5}{5}             & 5 / 5             & \RATIO{5}{5}             \\
            \ColHead{libxml2 (\cc{c})}                               & \PAD 4 / 6                                       & \RATIO{4}{6}                                                  & \PAD 5 / 6                               & \RATIO{5}{6}                                & \PAD 6 / 6                                     & \RATIO{6}{6}             & \PAD 6 / 6             & \RATIO{6}{6}             & \PAD 6 / 6             & \RATIO{6}{6}             & 6 / 6             & \RATIO{6}{6}             \\
            \ColHead{log4j2 (\cc{java})}                             & \PAD 1 / 1                                       & \RATIO{1}{1}                                                  & \PAD 1 / 1                               & \RATIO{1}{1}                                & \PAD 1 / 1                                     & \RATIO{1}{1}             & \PAD 1 / 1             & \RATIO{1}{1}             & \PAD 1 / 1             & \RATIO{1}{1}             & 1 / 1             & \RATIO{1}{1}             \\
            \ColHead{mongoose (\cc{c})}                              & \PAD 0 / 3                                       & \RATIO{0}{3}                                                  & \PAD 1 / 3                               & \RATIO{1}{3}                                & \PAD 2 / 3                                     & \RATIO{2}{3}             & \PAD 2 / 3             & \RATIO{2}{3}             & \PAD 2 / 3             & \RATIO{2}{3}             & 2 / 3             & \RATIO{2}{3}             \\
            \ColHead{pdfbox (\cc{java})}                             & \PAD 9 / 9                                       & \RATIO{9}{9}                                                  & \PAD 6 / 9                               & \RATIO{6}{9}                                & \PAD 8 / 9                                     & \RATIO{8}{9}             & \PAD 8 / 9             & \RATIO{8}{9}             & \PAD 9 / 9             & \RATIO{9}{9}             & 9 / 9             & \RATIO{9}{9}             \\
            \ColHead{shadowsocks (\cc{c})}                           & \PAD 5 / 5                                       & \RATIO{5}{5}                                                  & \PAD 5 / 5                               & \RATIO{5}{5}                                & \PAD 5 / 5                                     & \RATIO{5}{5}             & \PAD 5 / 5             & \RATIO{5}{5}             & \PAD 5 / 5             & \RATIO{5}{5}             & 5 / 5             & \RATIO{5}{5}             \\
            \ColHead{systemd (\cc{c})}                               & \PAD 3 / 4                                       & \RATIO{3}{4}                                                  & \PAD 4 / 4                               & \RATIO{4}{4}                                & \PAD 4 / 4                                     & \RATIO{4}{4}             & \PAD 4 / 4             & \RATIO{4}{4}             & \PAD 4 / 4             & \RATIO{4}{4}             & 4 / 4             & \RATIO{4}{4}             \\
            \ColHead{tika (\cc{java})}                               & \PAD 8 / 11                                      & \RATIO{8}{11}                                                 & \PAD 9 / 11                              & \RATIO{9}{11}                               & \PAD 10 / 11                                   & \RATIO{10}{11}           & \PAD 11 / 11           & \RATIO{11}{11}           & \PAD 9 / 11            & \RATIO{9}{11}            & 11 / 11           & \RATIO{11}{11}           \\
            \ColHead{wireshark (\cc{c})}                             & \PAD 12 / 12                                     & \RATIO{12}{12}                                                & \PAD 11 / 12                             & \RATIO{11}{12}                              & \PAD 12 / 12                                   & \RATIO{12}{12}           & \PAD 12 / 12           & \RATIO{12}{12}           & \PAD 12 / 12           & \RATIO{12}{12}           & 12 / 12           & \RATIO{12}{12}           \\
            \ColHead{xz (\cc{c})}                                    & \PAD 1 / 1                                       & \RATIO{1}{1}                                                  & \PAD 1 / 1                               & \RATIO{1}{1}                                & \PAD 1 / 1                                     & \RATIO{1}{1}             & \PAD 1 / 1             & \RATIO{1}{1}             & \PAD 1 / 1             & \RATIO{1}{1}             & 1 / 1             & \RATIO{1}{1}             \\
            \midrule
            \rowcolor{white}
            \RowHead{Totals}                                         & \PAD \RowHead{72 / 92}                           & \RowHead{\RATIO{72}{92}}                                      & \PAD \RowHead{76 / 92}                   & \RowHead{\RATIO{76}{92}}                    & \PAD \RowHead{84 / 92}                         & \RowHead{\RATIO{84}{92}} & \PAD \RowHead{84 / 92} & \RowHead{\RATIO{84}{92}} & \PAD \RowHead{83 / 92} & \RowHead{\RATIO{83}{92}} & \RowHead{87 / 92} & \RowHead{\RATIO{87}{92}} \\
            \bottomrule
        \end{tabular}
    \end{adjustbox}

\end{table*}

\subsection{\fp Orchestration}
\label{ss:rq_scheduling}

\begin{figure*}[t]
    \centering
    \begin{subfigure}[t]{0.45\textwidth}
        \centering
        \includegraphics[width=\linewidth]{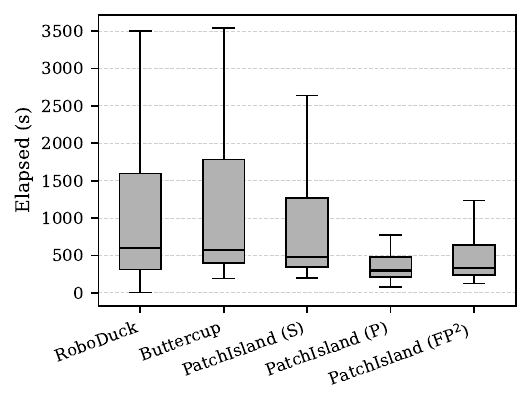}
        \caption{Time comparison across scheduling strategies and baselines.}
        \label{fig:time_boxplot}
    \end{subfigure}
    \hfill
    \begin{subfigure}[t]{0.45\textwidth}
        \centering
        \includegraphics[width=\linewidth]{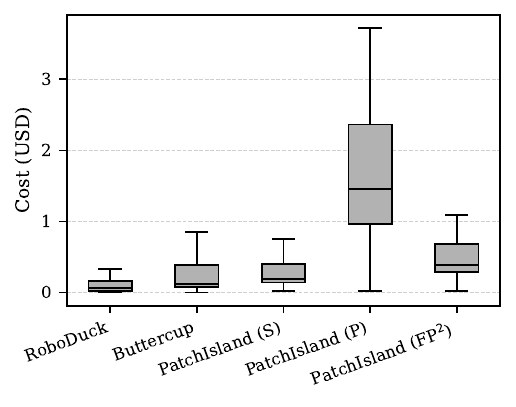}
        \caption{LLM inference cost across scheduling strategies and baselines.}
        \label{fig:cost_boxplot}
    \end{subfigure}
    \caption{
        Time and cost comparison of patch generation across scheduling strategies (Sequential~(S), Parallel~(P), and \fp) and baseline AIxCC systems.
        Each box shows the distribution of time and LLM inference cost required to produce successful patches, illustrating the latency (left) and LLM cost (right) trade-offs of different scheduling approaches.
    }
    \label{fig:fp_orchestration}
\end{figure*}

We now evaluate the effectiveness of \fp orchestration.
We compare three orchestration strategies:
(1) \emph{Sequential (S)}: agents run sequentially;
(2) \emph{Parallel (P)}: agents run in parallel; and
(3) \emph{FP$^2$}: agents run according to \fp orchestration. 

\PP{Balance between Execution Time and Cost}
As shown in \autoref{fig:fp_orchestration}, \sys achieves a good balance between execution time and LLM cost compared to baseline approaches.
These results are expected due to their inherent trade-offs.
Parallel execution minimizes latency by running all agents simultaneously, but incurs high costs as every agent participates in each attempt.
Sequential execution reduces cost by invoking only one agent at a time, but suffers from increased latency.
In contrast, \fp orchestration executes a small subset of agents in parallel and prioritizes high-performance agents to maximize efficiency.
Indeed, \fp orchestration allows the top-priority agent to resolve 91\% of successful cases, invoking more agents only for the remaining challenging cases.

\PP{Operational Robustness}
\fp orchestration can provide better operational robustness than other approaches.
Although it is difficult to evaluate with quantitative metrics,
we can reason about it.
For example, as DEI~\cite{dei},
if we use a parallel execution with centralized aggregation 
(e.g., LLM-as-a-judge to select the best patch),
the system can stop if one of the agents fails.
This is also true if we use a sequential execution.
Definitely, if we use a parallel execution with FCFS policy,
that might provide the highest robustness;
however, as shown in the previous section, it incurs a high cost.
Therefore, \fp orchestration could be a practical compromise between robustness and cost efficiency.

\section{Postmortem}
\label{s:post_mortem}

In this section, we present a postmortem analysis of our AIxCC final participation, demonstrating how \sys operates with real-world continuous fuzzing.

\PP{Summary}
The AIxCC final operated at a substantial scale.
It encompassed 53 Challenge Projects (CPs) derived from 28 repositories, containing approximately 54M lines of code and including 70 intended vulnerabilities.
During the final, each CP was evaluated for 6 hours in delta mode and 12 hours in full mode;
delta mode targets incremental changes, while full mode analyzes the entire codebase.
During the final, all teams collectively spent \$359K on compute resources and LLM usage.
After manual review, the organizers officially confirmed 43 distinct vulnerabilities and 31 distinct patches~\cite{aixcc-website}.
As shown in \autoref{tab:team_comparison}, we believe that these results are significant.  \sys
produced the largest number of patches among all participating teams (31) and
achieved the second-highest success rate (72.1\%). Notably, the team with the highest
success rate produced only one patch and achieved 100\% success rate.

\subsection{Methodology}
Our postmortem analysis relies on system logs and submission results.
After the competition, we could obtain submission results --- 122 Proofs of Vulnerability and 42 patches --- 64.7\,GB of system logs, and 192\,GB of artifacts, including CP tarballs.
Unfortunately, our data were incomplete due to technical issues;
We could not retain some logs and artifacts due to their large volume.
Nevertheless, these data remain informative and meaningful for understanding \sys’s performance in the realistic environment.
After the competition, the organizers provided a reference dataset, which was also used in \autoref{s:eval}.
This dataset included vulnerability labels and bug descriptions. However, unsurprisingly, our results did not include them.
Accordingly, we manually reconstructed the mapping by inspecting each CP. Through this process, we confirmed that our bug-finding system could generate PoVs for 31 out of the 53 CPs, for which \sys also attempted to generate patches. Accordingly, we restrict our analysis to these 31 CPs.

\begin{figure}[t]
    \centering
    \includegraphics[width=\linewidth, trim=0 2cm 0 0, clip]{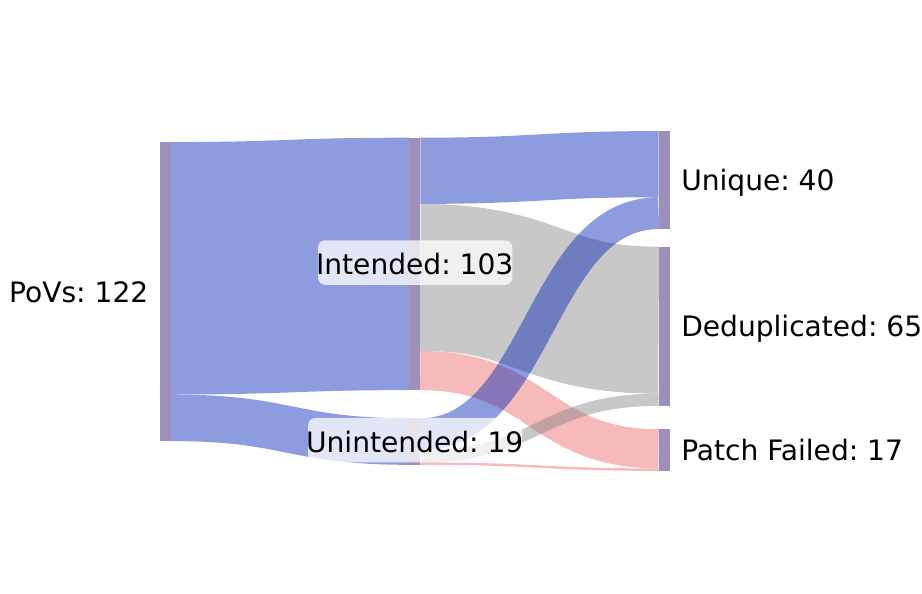}
    \caption{PoV deduplication based on successful patch generation. Among 99 intended PoVs, \sys successfully deduplicated them into 27 patches, while only two patches were generated for unintended PoVs.}
    \label{fig:dedup}
\end{figure}

\subsection{In-depth Analysis for Patching Results}
\label{ss:qanalysis}

We first conducted an in-depth analysis of patching results. This is motivated by the discrepancy between the number of patches submitted by \sys and that ultimately accepted by the AIxCC final. Although \sys submitted 42 patches, the organizers confirmed only 31 as correct in the final. Notably, in AIxCC, the organizers did not naively rely on plausibility; instead, they manually validated each patch to determine whether it correctly addressed the underlying vulnerability.

For that, we study how \sys handles intended and unintended vulnerabilities.
We define intended vulnerabilities as vulnerabilities that are deliberately injected by the organizers, each with an associated PoV in the reference dataset. In contrast, unintended vulnerabilities are the ones that the organizers did not intentionally introduce. These unintended vulnerabilities can be further categorized into two: (1) vulnerabilities that arise accidentally during the injection process, and (2) genuine zero-day vulnerabilities. Notably, our bug-finding system discovered two zero-day vulnerabilities during the final.
Moreover, we also analyze how effectively \sys performs deduplication in a continuous fuzzing environment.

\PP{Intended Vulnerability}
By re-running each PoV against the reference patch,
we could classify 99 out of 122 PoVs, including duplicates, into 33 distinct intended vulnerabilities.
Among these 33 vulnerabilities, \sys generated 28 plausible patches.
Our manual inspection confirmed that 27 of these patches are correct, while one is incorrect.
This yields a success rate of 82.8\% (27 / 33), which is consistent with our evaluation results (\autoref{s:eval}).
We found that \sys failed in the initialization phase of the \cc{systemd} challenge project,
despite our efforts on operational robustness. We will discuss this in more detail in \autoref{ss:case_studies}.
Although \sys generated correct 27 patches, these patches collectively covered 29 intended vulnerabilities
and two unintended vulnerabilities.
This is because one patch could fix three vulnerabilities at once, which we will discuss in detail in \autoref{ss:case_studies}.

\PP{Unintended Vulnerability}
For unintended vulnerabilities, \sys generated 14 patches: 4 correct patches and 10 plausible but incorrect ones.
Notably, one correct patch addresses a zero-day vulnerability, which we discuss further in \autoref{ss:realworldimpact}.
Compared to intended vulnerabilities, \sys produced more incorrect patches.
We believe this is due to the lack of functional tests for unintended vulnerabilities.
As a result, \sys readily generates plausible patches; however, these patches are often semantically incorrect.

\PP{Deduplication}
As shown in \autoref{fig:dedup}, the 42 patches deduplicated 65 PoVs out of the original 105 PoVs.
These 65 PoVs correspond to 53.3\% of the 122 total PoVs.
In addition, plausible but incorrect patches further reduced three PoVs; manual inspection confirmed that these cases were correctly deduplicated.
However, 17 PoVs arising from two vulnerabilities for which patch generation failed were not deduplicated.

\begin{table}[t]
    \centering
    \small
    \resizebox{\linewidth}{!}{%
        \begin{tabular}{l c r r r}
            \hline
            \textbf{Project Name} & \textbf{Scan Mode} & \textbf{Intended} & \textbf{Patched} & \textbf{Unpatched} \\
            \hline
            commons-compress      & Delta              & 4                     & 4                 & 0                   \\
            curl                  & Delta              & 2                     & 2                 & 0                   \\
            freerdp               & Delta              & 1                     & 0                 & 1                   \\
            libexif               & Delta              & 2                     & 2                 & 0                   \\
            libxml2               & Delta              & 1                     & 1                 & 0                   \\
            log4j2                & Delta              & 1                     & 1                 & 0                   \\
            mongoose              & Full               & 1                     & 1                 & 0                   \\
            mongoose              & Delta              & 2                     & 2                 & 0                   \\
            pdfbox                & Full               & 1                     & 1                 & 0                   \\
            pdfbox                & Delta              & 1                     & 0                 & 1                   \\
            shadowsocks           & Full               & 1                     & 1                 & 0                   \\
            systemd               & Full               & 3                     & 0                 & 3                   \\
            tika                  & Delta              & 1                     & 1                 & 0                   \\
            wireshark             & Full               & 5                     & 5                 & 0                   \\
            wireshark             & Delta              & 6                     & 5                 & 1                   \\
            xz                    & Full               & 1                     & 1                 & 0                   \\
            \hline
            \textbf{Total}        &                    & \textbf{33}           & \textbf{27}          & \textbf{6}            \\
            \hline
        \end{tabular}
    }
    \caption{Patch results for 33 distinct intended vulnerabilities. }
    \label{tab:intended_result}
\end{table}
\subsection{Case Studies}
\label{ss:case_studies}

In this section, we present two interesting case studies of \sys in the AIxCC final.

\begin{figure}[t]
    \centering
\begin{minted}{diff}
@@ -599,8 +599,8 @@ json_value * json_parse_ex (json_settings * settings,
 
 case 't': // PoV-triggered vulnerability

-   if ((end - state.ptr) < 3 || *(++ state.ptr) != 'r' ||
-       *(++ state.ptr) != 'u' || *(++ state.ptr) != 'e')
+  if ((end - state.ptr) < 4 || *(++ state.ptr) != 'r' ||
+      *(++ state.ptr) != 'u' || *(++ state.ptr) != 'e')
    {
       goto e_unknown_value;
    }
@@ -615,9 +615,9 @@ json_value * json_parse_ex (json_settings * settings,
 
 case 'f': // Undetected vulnerability

-   if ((end - state.ptr) < 4 || *(++ state.ptr) != 'a' ||
-       *(++ state.ptr) != 'l' || *(++ state.ptr) != 's' ||
-       *(++ state.ptr) != 'e')
+  if ((end - state.ptr) < 5 || *(++ state.ptr) != 'a' ||
+      *(++ state.ptr) != 'l' || *(++ state.ptr) != 's' ||
+      *(++ state.ptr) != 'e')
    {
       goto e_unknown_value;
    }
@@ -630,8 +630,8 @@ json_value * json_parse_ex (json_settings * settings,
 
 case 'n':  // Undetected vulnerability

-   if ((end - state.ptr) < 3 || *(++ state.ptr) != 'u' ||
-       *(++ state.ptr) != 'l' || *(++ state.ptr) != 'l')
+  if ((end - state.ptr) < 4 || *(++ state.ptr) != 'u' ||
+      *(++ state.ptr) != 'l' || *(++ state.ptr) != 'l')
    {
       goto e_unknown_value;
    }
\end{minted}

    \caption{A patch generated by \sys for a vulnerability in \cc{shadowsocks-libev}. The \cc{json_parse_ex()} function contains three distinct bugs across different \cc{case} branches, all of which are fixed simultaneously by this patch.}
    \label{fig:three_patch}
\end{figure}

\PP{One patch for three intended bugs}
During our analysis, we found that \sys generates a patch that fixes three intended vulnerabilities simultaneously, as shown in \autoref{fig:three_patch}.
It was out-of-bounds read vulnerability in \cc{json_parse_ex()} of \cc{shadowsocks-libev}.
Due to an incorrect check, the function may read beyond the end of the buffer when processing the keyword \cc{true}.
Interestingly, similar vulnerabilities exist for other keywords (i.e., \cc{false} and \cc{null}).
Since \sys treated these vulnerabilities as a single issue, it fixed all of them in a single patch.
As illustrated in this example, it is extremely subjective to define the equivalence of multiple vulnerabilities.
This also brings interesting issues in APR and its benchmarks.
We will discuss this in more detail in \autoref{s:discussion}.

\begin{figure}[t]
\begin{minted}{diff}
@@ -88,7 +88,13 @@ public class JsonSerializerDeserializer
   public Data deserialize(byte[] serializedJson) {
-    return gson.fromJson(new String(serializedJson), Data.class);
+    try {
+      return gson.fromJson(new String(serializedJson), Data.class);
+    } catch (ArrayIndexOutOfBoundsException e) {
+      return null;
+    }
   }
\end{minted}
    \caption{An example of an inadequate patch that suppresses failures through top-level exception handling instead of addressing the root cause inside \texttt{gson.fromJson()}. The patch prevents crashes by catching an exception, but leaves the underlying JSON parsing vulnerability unresolved.}
    \label{fig:bad_patch}
\end{figure}

\PP{Plausible but incorrect patch}
We also found that \sys could generate a plausible but incorrect patch.
\autoref{fig:bad_patch} illustrates an exmple of such patches;
Instead of addressing the root cause, \sys just suppressed the crash by introducing a top-level exception handler.
As expected, such cases occur more frequently for unintended vulnerabilities,
as they do not have any well-designed functional tests.
Specifically, we identified one (1/33) plausible but incorrect patch for intended vulnerabilities,
compared to (10/14) for unintended vulnerabilities.
These results underscore the importance of functional tests for evaluating AVR systems.

\subsection{Operational Robustness}
In the AIxCC final, \sys demonstrated its operational robustness thanks to its ensemble design. Through postmortem analysis, we identified multiple agent-level failures; however, these failures did not prevent successful patch generation. For instance, \sysmartian failed to generate patches because an LSP (Language Server Protocol) server failed during initialization. Nevertheless, \sys completed patch generation by relying on other agents that do not depend on LSP. Similarly, \sys remained effective under sporadic failures: although \sysmr failed due to LLM rate limiting, \sys still successfully generated patches.

\noindent \textbf{Broken symlink in \cc{systemd}.}
Nevertheless, \sys is not perfect. In practice, we found a critical failure in the \cc{systemd} challenge,
which exposed a single point of failure in \sys.
Notably, \sys assumes that the \coordinator has successfully initialized to operate.
To mitigate this risk, we deliberately designed the \coordinator to be as lightweight as possible;
however, we could not eliminate all issues.

In our case, the failure originated from broken symbolic links in \cc{systemd}. Specifically, \cc{systemd} contains a broken symlink
(i.e., \cc{etc/os-release} pointing to \cc{../usr/lib/os-release}, which can be invalid).
\sys includes logic to update anomalous timestamps, as they can make archives fail to be extracted.
However, when \sys checks timestamps for this broken symlink, it fails and causes the initialization to stop.
Due to this failure, \sys could not generate patches for three vulnerabilities.

This highlights the difficulty of achieving operational robustness in real-world settings.
Although we made substantial efforts to avoid such issues, practical deployments always expose
diverse and unexpected failures. Indeed, \teamc, \teame, and \teamf are also reported issues with operational robustness.
Nevertheless, we believe \sys's design remains meaningful as it could effectively mitigate agent-level failures, thereby substantially improving operational robustness.

\begin{figure}[t]
    \centering
\begin{minted}{diff}
--- a/pdfbox/src/main/java/org/apache/pdfbox/pdmodel/PDPageTree.java
+++ b/pdfbox/src/main/java/org/apache/pdfbox/pdmodel/PDPageTree.java
@@ -108,15 +108,28 @@
      */
     public static COSBase getInheritableAttribute(COSDictionary node, COSName key)
     {
+        return getInheritableAttribute(node, key, new HashSet<>());
+    }
+
+    private static COSBase getInheritableAttribute(COSDictionary node,
                                            COSName key, Set<COSDictionary> visited)
+    {
+        if (node == null || visited.contains(node))
+        {
+            return null;
+        }
+        
+        visited.add(node);
+        
         COSBase value = node.getDictionaryObject(key);
         if (value != null)
         {
             return value;
         }
+        
         COSDictionary parent = node.getCOSDictionary(COSName.PARENT, COSName.P);
         if (parent != null && COSName.PAGES.equals(parent.getCOSName(COSName.TYPE)))
         {
-            return getInheritableAttribute(parent, key);
+            return getInheritableAttribute(parent, key, visited);
         }
 
         return null;
\end{minted}
    \caption{The patch generated by \sys to fix infinite recursion in \cc{PDPageTree}. This diff matches the developer's commit, excluding one blank line.}

    \label{fig:pdfbox}
\end{figure}

\subsection{Real-World Impact}
\label{ss:realworldimpact}
We observed a \cc{pdfbox} 0-day bug in the unintended vulnerability analysis. This bug occurs when constructing the PDF Page Tree, where a loop can arise because previously visited attributes are not validated; \autoref{fig:pdfbox} shows the patch that prevents this behavior.
For this bug, the same patch as ours was suggested to the pdfbox maintainer and subsequently merged upstream~\cite{pdfbox_commit}. This suggestion and merge occurred during the competition evaluation period, after our system had generated the patch.
Although we cannot assert whether this patch originated from our system, this indicates that our research can contribute to the open-source community.

\begin{table}[t]
    \centering
    \small
    \caption{Comparison of performance with other teams on the AIxCC final.}
    \label{tab:team_comparison}
    \begin{adjustbox}{max width=\linewidth}
        \begin{tabular}{lccc}
            \toprule
            \textbf{System} &
            \makecell{\textbf{Number of }              \\ \textbf{Vulnerabilities}} &
            \makecell{\textbf{Number of }              \\ \textbf{Patches}} &
            \makecell{\textbf{Success}                 \\ \textbf{Rate}} \\
            \midrule
            \sys (Ours)     & 43 & 31 & \ratio{31}{43} \\
            \teamb          & 34 & 20 & \ratio{20}{34} \\
            \teama          & 28 & 19 & \ratio{19}{28} \\
            \teamd          & 28 & 14 & \ratio{14}{28} \\
            \teame          & 28 & 11 & \ratio{11}{28} \\
            \teamf          & 41 & 3  & \ratio{3}{41}  \\
            \teamc          & 1  & 1  & \ratio{1}{1}   \\

            \bottomrule
        \end{tabular}
    \end{adjustbox}
\end{table}

\section{Related Work}
\label{s:relwk}

\PP{LLM-based AVR}
Thanks to the recent advances in LLMs, many LLM-based AVR techniques have been proposed.
Among them, early approaches generate patches by directly prompting LLMs with bug reports and failing tests~\cite{pearce2023examining}.
More recent systems adopt agent-based designs with applicable tools, enabling iterative workflows that integrate code search, analysis, patch generation, and validation~\cite{aider,yang2024sweagent,autocoderover,agentless,Orwall_Moatless_Tools_2024,appatch,san2patch,repairagent}. 
Even under identical execution environments, AVR agents can exhibit substantially different performance.
Zhang \etal~\cite{dei} attribute this variation to inherent nondeterminism in LLM-based agents and to differences in agent design, including tools, workflows, and prompts.
Hu \etal~\cite{vuln4c_sok_avr} show that AVR techniques are not universally applicable across all environments due to their technical and environmental dependencies.

\PP{Ensemble-Based Approaches for AVR}
Recent studies explore ensemble-based approaches for automated repair, demonstrating that combining multiple repair agents or models can improve patching performance by increasing the diversity of candidate patches~\cite{eapr, papr, dei, mahmud2025enhancingllmcodegeneration}.
E-APR~\cite{eapr} selects APR tools using supervised machine learning, aiming to predict the most suitable repair tool for a given bug.
In contrast, P-EPR~\cite{papr} relies on manually defined repair patterns and preconditions to select APR tools, and re-ranks them based on feedback from previous repair attempts.
While these approaches focus on selecting a single most promising repair tool, other studies explicitly embrace redundancy by executing multiple agents in parallel despite the additional computational cost~\cite{dei}, or by generating multiple candidate patches and selecting the best one among them~\cite{mahmud2025enhancingllmcodegeneration}.
Our approach shares certain design aspects with P-EPR~\cite{papr}, in that both rely on manually defined criteria. 
However, in our preliminary evaluation, we did not observe substantial performance differences among LLM-based repair agents across different bug types. Consequently, we adopt alternative selection criteria based on agent characteristics and general performance rather than bug categories.

\PP{Crash Deduplication}
Existing fuzzers primarily deduplicate crashes based on execution-derived signals such as stack traces or behavioral features.
AFL++~\cite{afl} and ClusterFuzz~\cite{clusterfuzz} cluster crashes by hashing call stacks, under the assumption that crashes with similar stack traces correspond to the same underlying defect.
ReBucket~\cite{rebucket} clusters crashes using approximate stack traces, while Aurora~\cite{aurora} classifies failures by statistically distinguishing crashing and non-crashing behaviors.
Other approaches analyze call stacks using machine learning techniques to identify similar crashes~\cite{findingsimilar}.
Igor~\cite{igor} proposes a crash deduplication technique that is directly integrated into the fuzzing loop.
Despite their effectiveness in reducing redundant crash reports, these techniques typically remain decoupled from downstream repair stages, requiring an explicit handoff between crash deduplication and patch generation.

\section{Discussion}
\label{s:discussion}

\PP{Limitations}
Although \sys demonstrates its effectiveness in CVR, it has several limitations.
First, \sys exhibits a single point of failure during its initialization phase.
We attempted to mitigate this issue by minimizing the complexity of this phase.
This was effective; however, we still observed one failure during the AIxCC final.
Second, like other APR tools, \sys still can generate plausible yet incorrect patches (see \autoref{s:post_mortem}).
Finally, \sys incurs higher costs than existing baselines (see \autoref{s:eval}).
This stems from our design assumption; we assume that LLM credits are relatively abundant, as in the AIxCC competition setting.
For practicality, future work can explore ways to improve \sys's cost-effectiveness.

\PP{Metrics for AVR}
It remains an open problem to design effective metrics for AVR.
Our community commonly uses the number of plausible patches as a primary metric for evaluating AVR
~\cite{san2patch, appatch, csuvik2024genprogjs}.
However, as our postmortem analysis demonstrates, this metric is highly fragile.
First, the number of plausible patches can be artificially inflated.
As shown in \autoref{s:post_mortem}, a single vulnerability may be considered as multiple distinct vulnerabilities due to the lack of a clear definition of vulnerability equivalence.
Second, plausibility often fails to reflect the correctness, as mentioned in many prior studies
~\cite{liu2021critical, qi2015analysis, fei2025patch}.
Even though the AIxCC dataset attempts to carefully design its functional tests to mitigate this issue,
we still observed such cases in practice.
More seriously, our analysis indicates that this becomes more serious when we attempt to fix vulnerabilities without any functional tests
(e.g., unintended vulnerabilities).
To mitigate this, we manually inspected \sys's patches during our postmortem analysis; however,
this approach is not scalable and time-consuming for large-scale evaluation.

\section{Conclusion}
\label{s:conclusion}

In this paper, we presetned \sys, a continuous vulnerability repair (CVR) system integrated with continuous fuzzing.
\sys aims to be an effective, efficient, and operationally robust CVR system.
For that, \sys introduces several novel techniques, including ensemble-based approach, two-phase deduplication, and \fp orchestration.
Our evaluation on the AIxCC dataset demonstrated that \sys significantly outperforms the baseline systems, \baselinea and \baselineb, repairing 27 and 68 more vulnerabilities, respectively.
We further showed that \sys can generate correct patches for real-world vulnerabilities that can be even adopted by upstream projects.
Overall, \sys can be a promising direction to reduce manual efforts and improve scalability in real-world continuous fuzzing.

\bibliographystyle{ACM-Reference-Format}
\bibliography{p,sslab,conf}

@string{POT       = {Proceedings of the }}

@string{OSDI      = { USENIX Symposium on Operating Systems Design and Implementation (OSDI)}}

@string{FAST      = { USENIX Conference on File and Storage Technologies (FAST)}}

@string{SP        = { IEEE Symposium on Security and Privacy (Oakland)}}

@string{SEC       = { USENIX Security Symposium (Security)}}

@string{CCS       = { ACM Conference on Computer and Communications Security (CCS)}}

@string{WOOT      = { USENIX Workshop on Offensive Technologies (WOOT)}}

@string{ICSE      = { International Conference on Software Engineering (ICSE)}}

@string{FSE       = { ACM SIGSOFT Symposium on the Foundations of Software Engineering (FSE)}}

@string{ISSTA     = { International Symposium on Software Testing and Analysis (ISSTA)}}

@string{TOSEM     = { ACM Transactions on Software Engineering and Methodology (TOSEM)}}

@Proceedings{OSDI18,
  title        = POT # { 13th } # OSDI,
  booktitle    = POT # { 13th } # OSDI,
  month        = oct,
  year         = 2018,
  address      = {Carlsbad, CA},
}

@Proceedings{CCS21,
  title        = POT # { 28th } # CCS,
  booktitle    = POT # { 28th } # CCS,
  month        = nov,
  year         = 2021,
  address      = {Virtual},
}

@Proceedings{SEC20,
  title       = POT # { 29th } # SEC,
  booktitle   = POT # { 29th } # SEC,
  month       = aug,
  year        = 2020,
  address     = {Boston, MA},
}

@Proceedings{SEC25,
  title       = POT # { 34th } # SEC,
  booktitle   = POT # { 34th } # SEC,
  month       = aug,
  year        = 2025,
  address     = {Seattle, WA},
}

@Proceedings{SP19,
  title       = POT # { 40th } # SP,
  booktitle   = POT # { 40th } # SP,
  month       = may,
  year        = 2019,
  address     = {San Francisco, CA},
}

@Proceedings{ICSE12,
  title        = POT # { 34th } # ICSE,
  booktitle    = POT # { 34th } # ICSE,
  month        = jun,
  year         = 2012,
  address      = {Zurich, Switzerland},
}

@Proceedings{ICSE16,
  title        = POT # { 38th } # ICSE,
  booktitle    = POT # { 38th } # ICSE,
  month        = may,
  year         = 2016,
  address      = {Texas, USA},
}

@Proceedings{ICSE24,
  title        = POT # { 46th } # ICSE,
  booktitle    = POT # { 46th } # ICSE,
  month        = apr,
  year         = 2024,
  address      = {Lisbon, Portugal},
}

@Proceedings{ICSE25,
  title        = POT # { 47th } # ICSE,
  booktitle    = POT # { 47th } # ICSE,
  month        = sep,
  year         = 2025,
  address      = {Ottawa, Canada},
}

@proceedings{ICSE26,
  title        = POT # { 48th } # ICSE,
  booktitle    = POT # { 48th } # ICSE,
  month        = apr,
  year         = 2026,
  address      = {Rio, Brazil},
}

@Proceedings{TOSEM21,
  title        = POT # { } # TOSEM,
  booktitle    = POT # { } # TOSEM,
  month        = feb,
  year         = 2021,
  address      = {New York, NY, USA},
}

@Proceedings{WOOT20,
  title        = POT # { 14th } # WOOT,
  booktitle    = POT # { 14th } # WOOT,
  month        = aug,
  year         = 2020,
  address      = {Boston, MA},
}

@inproceedings{autocoderover,
author = {Zhang, Yuntong and Ruan, Haifeng and Fan, Zhiyu and Roychoudhury, Abhik},
title = {AutoCodeRover: Autonomous Program Improvement},
year = {2024},
isbn = {9798400706127},
publisher = {Association for Computing Machinery},
address = {New York, NY, USA},
url = {https://doi.org/10.1145/3650212.3680384},
doi = {10.1145/3650212.3680384},
booktitle = {Proceedings of the 33rd ACM SIGSOFT International Symposium on Software Testing and Analysis},
pages = {1592–1604},
numpages = {13},
location = {Vienna, Austria},
series = {ISSTA 2024}
}

@misc{aider,
  author = {Aider-AI},
  title = {aider: AI pair programming in your terminal},
  url = {https://github.com/Aider-AI/aider},
  note = {Accessed: Nov. 6, 2025}
}

@inproceedings{yang2024sweagent,
    title={{SWE}-agent: Agent-Computer Interfaces Enable Automated Software Engineering},
    author={John Yang and Carlos E Jimenez and Alexander Wettig and Kilian Lieret and Shunyu Yao and Karthik R Narasimhan and Ofir Press},
    booktitle={The Thirty-eighth Annual Conference on Neural Information Processing Systems},
    year={2024},
    url={https://arxiv.org/abs/2405.15793}
}

@article{agentless,
author = {Xia, Chunqiu Steven and Deng, Yinlin and Dunn, Soren and Zhang, Lingming},
title = {Demystifying LLM-Based Software Engineering Agents},
year = {2025},
issue_date = {July 2025},
publisher = {Association for Computing Machinery},
address = {New York, NY, USA},
volume = {2},
number = {FSE},
url = {https://doi.org/10.1145/3715754},
doi = {10.1145/3715754},
journal = {Proc. ACM Softw. Eng.},
month = jun,
articleno = {FSE037},
numpages = {24}
}

@misc{Orwall_Moatless_Tools_2024,
  author = {Örwall, Albert},
  title = {{Moatless Tools}},
  year = {2024},
  doi = {10.5281/zenodo.15614422},
  url = {https://github.com/aorwall/moatless-tools},
  note = {License: MIT}
}

@inproceedings{appatch,
author = {Nong, Yu and Yang, Haoran and Cheng, Long and Hu, Hongxin and Cai, Haipeng},
title = {APPATCH: automated adaptive prompting large language models for real-world software vulnerability patching},
crossref = {SEC25}
}

@inproceedings{san2patch,
  author = {Kim, Youngjoon and Shin, Sunguk and Kim, Hyoungshick and Yoon, Jiwon},
  title = {Logs in, patches out: automated vulnerability repair via tree-of-thought LLM analysis},
  crossref = {SEC25}
}

@inproceedings{pearce2023examining,
  title={Examining zero-shot vulnerability repair with large language models},
  author={Pearce, Hammond and Tan, Benjamin and Ahmad, Baleegh and Karri, Ramesh and Dolan-Gavitt, Brendan},
  booktitle={2023 IEEE Symposium on Security and Privacy (SP)},
  pages={2339--2356},
  year={2023},
  organization={IEEE}
}

@inproceedings{repairagent,
  author = {Bouzenia, Islem and Devanbu, Premkumar and Pradel, Michael},
  title = {{RepairAgent}: An Autonomous, LLM-Based Agent for Program Repair},
  crossref = {ICSE25}
}

@misc{zhang2024fixingsecurityvulnerabilitiesai,
      title={Fixing Security Vulnerabilities with AI in OSS-Fuzz}, 
      author={Yuntong Zhang and Jiawei Wang and Dominic Berzin and Martin Mirchev and Dongge Liu and Abhishek Arya and Oliver Chang and Abhik Roychoudhury},
      year={2024},
      eprint={2411.03346},
      archivePrefix={arXiv},
      primaryClass={cs.CR},
      url={https://arxiv.org/abs/2411.03346}, 
}

@ARTICLE{genprog,
  author={Le Goues, Claire and Nguyen, ThanhVu and Forrest, Stephanie and Weimer, Westley},
  journal={IEEE Transactions on Software Engineering}, 
  title={GenProg: A Generic Method for Automatic Software Repair}, 
  year={2012}
  }

@inproceedings{senx,
  author={Huang, Zhen and Lie, David and Tan, Gang and Jaeger, Trent},
  title={Using Safety Properties to Generate Vulnerability Patches}, 
  crossref = {SP19}
}

@inproceedings{extractfix,
  author = {Gao, Xiang and Wang, Bo and Duck, Gregory J. and Ji, Ruyi and Xiong, Yingfei and Roychoudhury, Abhik},
  title = {Beyond Tests: Program Vulnerability Repair via Crash Constraint Extraction},
  crossref = {TOSEM21}
}

@misc{codex_old,
      title={Evaluating Large Language Models Trained on Code}, 
      author={Mark Chen and Jerry Tworek and Heewoo Jun and Qiming Yuan and Henrique Ponde de Oliveira Pinto and Jared Kaplan and Harri Edwards and Yuri Burda and Nicholas Joseph and Greg Brockman and Alex Ray and Raul Puri and Gretchen Krueger and Michael Petrov and Heidy Khlaaf and Girish Sastry and Pamela Mishkin and Brooke Chan and Scott Gray and Nick Ryder and Mikhail Pavlov and Alethea Power and Lukasz Kaiser and Mohammad Bavarian and Clemens Winter and Philippe Tillet and Felipe Petroski Such and Dave Cummings and Matthias Plappert and Fotios Chantzis and Elizabeth Barnes and Ariel Herbert-Voss and William Hebgen Guss and Alex Nichol and Alex Paino and Nikolas Tezak and Jie Tang and Igor Babuschkin and Suchir Balaji and Shantanu Jain and William Saunders and Christopher Hesse and Andrew N. Carr and Jan Leike and Josh Achiam and Vedant Misra and Evan Morikawa and Alec Radford and Matthew Knight and Miles Brundage and Mira Murati and Katie Mayer and Peter Welinder and Bob McGrew and Dario Amodei and Sam McCandlish and Ilya Sutskever and Wojciech Zaremba},
      year={2021},
      eprint={2107.03374},
      archivePrefix={arXiv},
      primaryClass={cs.LG},
      url={https://arxiv.org/abs/2107.03374}
}

@inproceedings{AlphaRepair,
  author = {Xia, Chunqiu Steven and Zhang, Lingming},
  title = {Less training, more repairing please: revisiting automated program repair via zero-shot learning},
  year = {2022},
  isbn = {9781450394130},
  publisher = {Association for Computing Machinery},
  address = {New York, NY, USA},
  url = {https://doi.org/10.1145/3540250.3549101},
  doi = {10.1145/3540250.3549101},
  booktitle = {Proceedings of the 30th ACM Joint European Software Engineering Conference and Symposium on the Foundations of Software Engineering},
  pages = {959-971},
  numpages = {13},
  location = {Singapore, Singapore},
  series = {ESEC/FSE 2022}
}

@article{csuvik2024genprogjs,
  title={GenProgJS: A Baseline System for Test-Based Automated Repair of JavaScript Programs},
  author={Csuvik, Viktor and Horv{\'a}th, D{\'a}niel and Lajk{\'o}, M{\'a}rk and Vid{\'a}cs, L{\'a}szl{\'o}},
  journal={IEEE Transactions on Software Engineering},
  number={01},
  pages={1--19},
  year={2024},
  publisher={IEEE Computer Society}
}

@article{liu2021critical,
  title={A critical review on the evaluation of automated program repair systems},
  author={Liu, Kui and Li, Li and Koyuncu, Anil and Kim, Dongsun and Liu, Zhe and Klein, Jacques and Bissyand{\'e}, Tegawend{\'e} F},
  journal={Journal of Systems and Software},
  volume={171},
  pages={110817},
  year={2021},
  publisher={Elsevier}
}

@inproceedings{qi2015analysis,
  title={An analysis of patch plausibility and correctness for generate-and-validate patch generation systems},
  author={Qi, Zichao and Long, Fan and Achour, Sara and Rinard, Martin},
  booktitle={Proceedings of the 2015 international symposium on software testing and analysis},
  pages={24--36},
  year={2015}
}

@article{fei2025patch,
  title={Patch correctness assessment: A survey},
  author={Fei, Zhiwei and Ge, Jidong and Li, Chuanyi and Wang, Tianqi and Li, Yuning and Zhang, Haodong and Huang, LiGuo and Luo, Bin},
  journal={ACM Transactions on Software Engineering and Methodology},
  volume={34},
  number={2},
  pages={1--50},
  year={2025},
  publisher={ACM New York, NY}
}

@inproceedings{vuln4c_sok_avr,
  author = {Yiwei Hu and Zhen Li and Kedie Shu and Shenghua Guan and Deqing Zou and Shouhuai Xu and Bin Yuan and Hai Jin},
  title = {{SoK}: automated vulnerability repair: methods, tools, and assessments},
  crossref = {SEC25},
}

@misc{dei,
    title={Diversity Empowers Intelligence: Integrating Expertise of Software Engineering Agents}, 
    author={Kexun Zhang and Weiran Yao and Zuxin Liu and Yihao Feng and Zhiwei Liu and Rithesh Murthy and Tian Lan and Lei Li and Renze Lou and Jiacheng Xu and Bo Pang and Yingbo Zhou and Shelby Heinecke and Silvio Savarese and Huan Wang and Caiming Xiong},
    year={2024},
    eprint={2408.07060},
    archivePrefix={arXiv},
    primaryClass={cs.SE},
    url={https://arxiv.org/abs/2408.07060}, 
}

@inproceedings{papr,
  author = {Zhong, Wenkang and Li, Chuanyi and Liu, Kui and Xu, Tongtong and Ge, Jidong and Bissyande, Tegawende F. and Luo, Bin and Ng, Vincent},
  title = {Practical Program Repair via Preference-based Ensemble Strategy},
  crossref = {ICSE24}
}

@misc{mahmud2025enhancingllmcodegeneration,
      title={Enhancing LLM Code Generation with Ensembles: A Similarity-Based Selection Approach}, 
      author={Mahmud, Tarek and Duan, Bin and Pasareanu, Corina and Yang, Guowei},
      year={2025},
      eprint={2503.15838},
      archivePrefix={arXiv},
      primaryClass={cs.SE},
      url={https://arxiv.org/abs/2503.15838}, 
}

@article{eapr,
  author = {Aleti, Aldeida and Martinez, Matias},
  title = {E-APR: Mapping the effectiveness of automated program repair techniques},
  year = {2021},
  issue_date = {Sep 2021},
  publisher = {Kluwer Academic Publishers},
  address = {USA},
  volume = {26},
  number = {5},
  issn = {1382-3256},
  url = {https://doi.org/10.1007/s10664-021-09989-x},
  doi = {10.1007/s10664-021-09989-x},
  journal = {Empirical Softw. Engg.},
  month = sep,
  numpages = {30},
}

@inproceedings{igor,
  author = {Jiang, Zhiyuan and Jiang, Xiyue and Hazimeh, Ahmad and Tang, Chaojing and Zhang, Chao and Payer, Mathias},
  title = {Igor: Crash Deduplication Through Root-Cause Clustering},
  crossref = {CCS21}
}

@inproceedings {afl,
  author = {Andrea Fioraldi and Dominik Maier and Heiko Ei{\ss}feldt and Marc Heuse},
  title = {{AFL++ : Combining Incremental Steps of Fuzzing Research}},
  crossref = {WOOT20}
}

@misc{clusterfuzz,
  author = {{Google}},
  title = {{ClusterFuzz}},
  url = {https://google.github.io/clusterfuzz/},
  note = {Accessed: 2025-12-01}
}

@misc{honggfuzz, 
  author = {Google},
  title = {honggfuzz: General-purpose, easy-to-use fuzzer with interesting analysis options},
  url = {https://github.com/google/honggfuzz},
  note = {Accessed: 2025-12-01}
}

@inproceedings{GPTrace,
  title = {GPTrace: Effective Crash Deduplication Using LLM Embeddings},
  author = {Herter, Patrick and Ahlrichs, Vincent and A{\c{c}}ilan, Ridvan and Horsch, Julian},
  crossref = {ICSE26}
}

@inproceedings{RETracer,
  author={Cui, Weidong and Peinado, Marcus and Cha, Sang Kil and Fratantonio, Yanick and Kemerlis, Vasileios P.},
  title={RETracer: Triaging Crashes by Reverse Execution from Partial Memory Dumps}, 
  crossref = {ICSE16}
}

@inproceedings{rept,
  author = {Weidong Cui and Xinyang Ge and Baris Kasikci and Ben Niu and Upamanyu Sharma and Ruoyu Wang and Insu Yun},
  title = {{REPT}: Reverse Debugging of Failures in Deployed Software},
  crossref = {OSDI18}
}

@inproceedings{aurora,
  author = {Tim Blazytko and Moritz Schl{\"o}gel and Cornelius Aschermann and Ali Abbasi and Joel Frank and Simon W{\"o}rner and Thorsten Holz},
  title = {AURORA: Statistical Crash Analysis for Automated Root Cause Explanation},
  crossref = {SEC20}
}

@inproceedings{findingsimilar,
  author = {Bartz, Kevin and Stokes, Jack W. and Platt, John C. and Kivett, Ryan and Grant, David and Calinoiu, Silviu and Loihle, Gretchen},
  title = {Finding similar failures using callstack similarity},
  year = {2008},
  publisher = {USENIX Association},
  address = {USA},
  booktitle = {Proceedings of the Third Conference on Tackling Computer Systems Problems with Machine Learning Techniques},
  pages = {1},
  numpages = {1},
  location = {San Diego, California},
  series = {SysML'08}
}

@inproceedings{rebucket,
  author = {Dang, Yingnong and Wu, Rongxin and Zhang, Hongyu and Zhang, Dongmei and Nobel, Peter},
  title = {ReBucket: a method for clustering duplicate crash reports based on call stack similarity},
  crossref = {ICSE12}
}

@inproceedings{PeachCI,
  author = {Chen, Wanli and Chen, Yuanliang and Ma, Fuchen and Peng, Ruikang and Xu, Qi and Jiang, Yu and Fu, Qiang and Shi, Heyuan},
  title = {PeachCI: Scalable Continuous Integration of Generation-Based Protocol Fuzzing},
  year = {2025},
  isbn = {9798400712760},
  publisher = {Association for Computing Machinery},
  address = {New York, NY, USA},
  booktitle = {Proceedings of the 33rd ACM International Conference on the Foundations of Software Engineering (FSE Companion '25)},
  pages = {1213-1217},
  numpages = {5}
}

@inproceedings{EffectiveFuzzingCI,
author = {Sharma, Arindam and Cadar, Cristian and Metzman, Jonathan},
title = {Effective Fuzzing within CI/CD Pipelines (Registered Report)},
year = {2024},
isbn = {9798400711121},
publisher = {Association for Computing Machinery},
address = {New York, NY, USA},
url = {https://doi.org/10.1145/3678722.3685534},
doi = {10.1145/3678722.3685534},
booktitle = {Proceedings of the 3rd ACM International Fuzzing Workshop},
pages = {52-60},
numpages = {9},
location = {Vienna, Austria},
series = {FUZZING 2024}
}

@misc{CWE-193,
  author = {{The MITRE Corporation}},
  title = {{CWE-193: Off-by-one Error}},
  url = {https://cwe.mitre.org/data/definitions/193.html},
  note = {Accessed: 2025-12-01}
}

@misc{ossfuzz,
  author = {{Google}},
  title = {{OSS-Fuzz}: Continuous Fuzzing for Open Source Software},
  year = {2025},
  url = {https://google.github.io/oss-fuzz/},
  note = {Accessed: 2025-12-01}
}

@manual{intel_pt,
  title        = {Intel 64 and IA-32 Architectures Software Developer's Manual},
  author       = {{Intel Corporation}},
  year         = {2025},
  url          = {https://www.intel.com/content/www/us/en/developer/articles/technical/intel-sdm.html}
}

@misc{aixcc-website,
  author = {{DARPA}},
  title = {{AI Cyber Challenge (AIxCC)}},
  year = {2024},
  url = {https://aicyberchallenge.com/},
  note = {Accessed: 2025-11-07}
}

@misc{aixcc-scoring,
  author = {{DARPA}},
  title = {{Final Competition Procedures and Scoring Guide}},
  year = {2025},
  url = {https://aicyberchallenge.com/final-competition-procedures-and-scoring-guide/},
  note = {Accessed: 2025-11-07}
}

@misc{roboduck,
  author = {{Theori}},
  title = {{Robo Duck}},
  url = {https://github.com/theori-io/aixcc-afc-archive},
  note = {Accessed: 2025-12-01}}

@misc{buttercup,
  author = {{Trail of Bits}},
  title = {{Buttercup}},
  url = {https://github.com/trailofbits/buttercup},
  note = {Accessed: 2025-12-01}}

@misc{ctags,
  author       = {Universal Ctags},
  title        = {ctags},
  url = {https://github.com/universal-ctags/ctags},
  note         = {Accessed: 2025-07-27},
  year         = {2025}
}

@misc{jstack,
  author       = {{Oracle}},
  title        = {{jstack - Stack Trace}},
  url = {https://docs.oracle.com/javase/7/docs/technotes/tools/share/jstack.html},
}

@misc{techreport,
  author = {Team-Atlanta},
  title = {Technical Report},
  url = {https://arxiv.org/abs/2509.14589},
  year = {2025},
}

@misc{clangd,
  title={{clangd}},
  author={{clangd}},
  howpublished={\url{https://clangd.llvm.org}},
  note={Accessed: 2025-12-10}
}

@misc{eclipsejdtls,
  title={{Eclipse JDT Language Server}},
  author={{eclipse-jdtls}},
  howpublished={\url{https://github.com/eclipse-jdtls/eclipse.jdt.ls}},
  note={Accessed: 2025-12-10}
}

@misc{ccache,
  author = {ccache},
  title = {{Ccache - a fast C/C++ compiler cache}},
  url = {https://ccache.dev/},
  note={Accessed: 2025-12-10}
}

@misc{treesitter,
  author = {Max Brunsfeld},
  title = {{Tree-sitter}},
  url = {https://tree-sitter.github.io/tree-sitter/},
  note={Accessed: 2025-12-10}
}

@misc{pdfbox_commit,
  author = {Apache},
  url = {https://github.com/apache/pdfbox/commit/e58a1c92e97f9a207aaa5df5777d1ee6edb1104c},
  year = {2025},
  note = {Accessed: 2025-12-10}
}

\appendix

\newpage
\onecolumn
{\small
\begin{longtable}{l l p{0.6\linewidth}}
\caption{%
Detailed list of CP vulnerabilities included in our appendix dataset.
This table enumerates all 92 vulnerabilities selected from the AIxCC CP dataset
that include an Intended Proof-of-Vulnerability (PoV), categorized by project,
dataset mode (delta or full), and vulnerability type.
}
\label{tab:appendix-cp-vulns} \\

\toprule
\textbf{Project Name} & \textbf{Mode} & \textbf{Bug Type} \\
\midrule
\endfirsthead

\toprule
\textbf{Project Name} & \textbf{Mode} & \textbf{Bug Type} \\
\midrule
\endhead

\midrule
\multicolumn{3}{r}{\emph{Continued on next page}} \\
\endfoot

\bottomrule
\endlastfoot
apache-commons-compress & delta &
\detokenize{Regular expression denial of service (ReDoS) / catastrophic backtracking leading to StackOverflow} \\

apache-commons-compress & delta &
\detokenize{Uncontrolled memory allocation (OOM)} \\

apache-commons-compress & delta &
\detokenize{Denial of Service (Uncontrolled resource consumption via BigDecimal high-precision parsing)} \\

apache-commons-compress & delta &
\detokenize{Path traversal (Zip Slip via URL-encoded paths)} \\

apache-commons-compress & delta &
\detokenize{OS command injection} \\

apache-commons-compress & delta &
\detokenize{Zip Slip (path traversal via Unicode normalization)} \\

apache-commons-compress & delta &
\detokenize{Path traversal (Zip Slip via symlinks)} \\

apache-commons-compress & full &
\detokenize{Server-Side Request Forgery (SSRF)} \\

apache-commons-compress & full &
\detokenize{Command injection (backdoor / arbitrary command execution)} \\

apache-commons-compress & full &
\detokenize{Insecure deserialization (CWE-502)} \\

apache-commons-compress & full &
\detokenize{Zip Slip / Path Traversal} \\

apache-poi & delta &
\detokenize{Server-Side Request Forgery (SSRF)} \\

apache-poi & delta &
\detokenize{Regular expression denial of service (ReDoS) causing StackOverflowError} \\

apache-poi & full &
\detokenize{Unbounded loop / excessive iteration (resource exhaustion / DoS)} \\

apache-poi & full &
\detokenize{Uncontrolled memory allocation (memory allocation with excessive size)} \\

apache-poi & full &
\detokenize{Path traversal (directory traversal)} \\

apache-poi & full &
\detokenize{Logic bomb / backdoor via improper use of Unsafe (crash triggered by crafted sheet name)} \\

apache-poi & full &
\detokenize{Backdoor-triggered unexpected termination (unsafe use of Runtime.halt/System.exit; CWE-382)} \\

curl & delta & \detokenize{Null pointer dereference} \\
curl & delta & \detokenize{Null pointer dereference} \\
curl & delta & \detokenize{Null pointer dereference} \\
curl & delta & \detokenize{Stack buffer overflow} \\
curl & delta & \detokenize{Null pointer dereference} \\
curl & delta & \detokenize{Format string vulnerability} \\
curl & delta & \detokenize{Out-of-bounds read} \\

curl & full &
\detokenize{Out-of-bounds write (buffer overflow)} \\

dav1d & full &
\detokenize{Integer overflow} \\

freerdp & delta &
\detokenize{Integer overflow leading to heap-based buffer overflow} \\

freerdp & delta &
\detokenize{Heap-based buffer overflow} \\

freerdp & delta &
\detokenize{Write-what-where (arbitrary memory write)} \\

freerdp & full &
\detokenize{Integer overflow (signedness) leading to heap buffer overflow} \\

freerdp & full &
\detokenize{Code injection (obfuscated backdoor enabling arbitrary code execution)} \\

lcms & full &
\detokenize{Null pointer dereference} \\

lcms & full &
\detokenize{Buffer over-read} \\

libavif & delta &
\detokenize{Buffer overflow (out-of-bounds write)} \\

libexif & delta &
\detokenize{Heap-based buffer overflow} \\

libexif & delta &
\detokenize{Buffer over-read (out-of-bounds read)} \\

libexif & delta &
\detokenize{Heap buffer overflow} \\

libexif & delta &
\detokenize{Heap buffer overflow (out-of-bounds write)} \\

libexif & delta &
\detokenize{Buffer overflow} \\

libxml2 & delta &
\detokenize{Heap-based buffer overflow} \\

libxml2 & delta &
\detokenize{Double free} \\

libxml2 & delta &
\detokenize{Heap-based buffer overflow} \\

libxml2 & delta &
\detokenize{Heap-based buffer overflow} \\

libxml2 & full &
\detokenize{Heap-based buffer overflow} \\

libxml2 & full &
\detokenize{Use-after-free} \\

log4j2 & delta &
\detokenize{JNDI injection (remote code execution via Log4Shell)} \\

mongoose & delta &
\detokenize{Stack-based buffer overflow} \\

mongoose & delta &
\detokenize{Out-of-bounds read (read buffer overflow)} \\

mongoose & full &
\detokenize{Stack-based buffer overflow (off-by-one)} \\

pdfbox & delta &
\detokenize{Command injection} \\

pdfbox & full &
\detokenize{XML External Entity (XXE) leading to SSRF} \\

pdfbox & full &
\detokenize{XML External Entity (XXE) leading to SSRF} \\

pdfbox & full &
\detokenize{Infinite loop} \\

pdfbox & full &
\detokenize{Infinite loop} \\

pdfbox & full &
\detokenize{Uncontrolled memory allocation (CWE-789)} \\

pdfbox & full &
\detokenize{Integer overflow} \\

pdfbox & full &
\detokenize{Uncontrolled recursion / stack overflow (missing cycle detection in page tree traversal)} \\

pdfbox & full &
\detokenize{Infinite loop (CWE-834: excessive iteration due to circular xref references)} \\

shadowsocks & full &
\detokenize{Buffer over-read (heap-based, off-by-one)} \\

shadowsocks & full &
\detokenize{Buffer over-read (heap-based)} \\

shadowsocks & full &
\detokenize{Heap buffer over-read (CWE-126), off-by-one} \\

shadowsocks & full &
\detokenize{Buffer over-read (CWE-126)} \\

shadowsocks & full &
\detokenize{Buffer over-read (heap-based, off-by-one)} \\

systemd & full &
\detokenize{Out-of-bounds write (buffer overflow)} \\

systemd & full &
\detokenize{Heap-based buffer overflow (out-of-bounds write)} \\

systemd & full &
\detokenize{Double free} \\

systemd & full &
\detokenize{Double-free} \\

tika & delta &
\detokenize{OS command injection (CWE-78)} \\

tika & delta &
\detokenize{Server-Side Request Forgery (SSRF)} \\

tika & delta &
\detokenize{XML External Entity (XXE) vulnerability} \\

tika & delta &
\detokenize{Path traversal (Zip Slip)} \\

tika & delta &
\detokenize{Path traversal (Zip Slip)} \\

tika & delta &
\detokenize{Algorithmic complexity vulnerability (excessive iteration / performance DoS)} \\

tika & full &
\detokenize{XML External Entity (XXE)} \\

tika & full &
\detokenize{Insecure object deserialization} \\

tika & full &
\detokenize{Path traversal (Zip Slip) in archive extraction} \\

tika & full &
\detokenize{Server-Side Request Forgery (SSRF)} \\

tika & full &
\detokenize{OS command injection} \\

wireshark & delta &
\detokenize{Buffer underwrite (out-of-bounds write)} \\

wireshark & delta &
\detokenize{Stack-based buffer overflow} \\

wireshark & delta &
\detokenize{Heap-based buffer overflow} \\

wireshark & delta &
\detokenize{Buffer over-read (heap out-of-bounds read)} \\

wireshark & delta &
\detokenize{Out-of-bounds array index (improper index validation)} \\

wireshark & delta &
\detokenize{Buffer overflow (global buffer overflow)} \\

wireshark & full &
\detokenize{Buffer overflow (stack-based)} \\

wireshark & full &
\detokenize{Use-after-free} \\

wireshark & full &
\detokenize{Format string vulnerability (uncontrolled format string)} \\

wireshark & full &
\detokenize{Heap-based buffer overflow} \\

wireshark & full &
\detokenize{Heap-based buffer overflow} \\

wireshark & full &
\detokenize{Use-after-free} \\

xz & full &
\detokenize{Use-after-free} \\

\end{longtable}
}
\newpage

\begin{table*}[h]
    \centering
    \caption{Evaluation results for five agents over 92 AIxCC vulnerability instances. Each entry indicates whether the agent succeeded. The rightmost column reflects the maximum performance of \sys, serving as an upper-bound baseline in which any agent success counts.}
    \rowcolors{2}{gray!20}{white}

    \newcommand{\RowHead}[1]{\textbf{\makecell{\rule{0pt}{1.5em}#1\rule[-1.3em]{0pt}{0pt}}}}
    \newcommand{\ColHead}[1]{\textbf{\makecell{#1}}}
    \resizebox{!}{9.3cm} {%
        \begin{tabular}{
            |p{10cm}|C{2cm}C{2cm}C{2cm}C{2cm}C{2cm}||C{2.3cm}|
            }
            \hline
            \RowHead{\rule{0pt}{2em}Vulnerability ID}                      &
            \RowHead{\sysclaudelike}                                       &
            \RowHead{\sysmartian}                                          &
            \RowHead{\sysmr}                                               &
            \RowHead{\sysprism}                                            &
            \RowHead{\sysvincent}                                          &
            \RowHead{\sys}                                                                                                                                                                                                               \\
            \hline
            \ColHead{apache-commons-compress-cc-delta-01-vuln\_3}          & \BAD                      & \SOUND                    & \SOUND                    & \SOUND                    & \SOUND                    & \PASS                     \\
            \ColHead{apache-commons-compress-cc-delta-02-vuln\_9}          & \SOUND                    & \SOUND                    & \SOUND                    & \SOUND                    & \BAD                      & \PASS                     \\
            \ColHead{apache-commons-compress-cc-delta-03-vuln\_4}          & \SOUND                    & \SOUND                    & \SOUND                    & \SOUND                    & \SOUND                    & \PASS                     \\
            \ColHead{apache-commons-compress-cc-delta-04-vuln\_6}          & \SOUND                    & \SOUND                    & \SOUND                    & \SOUND                    & \SOUND                    & \PASS                     \\
            \ColHead{apache-commons-compress-cc-delta-05-vuln\_10}         & \SOUND                    & \SOUND                    & \SOUND                    & \SOUND                    & \SOUND                    & \PASS                     \\
            \ColHead{apache-commons-compress-cc-delta-06-vuln\_7}          & \SOUND                    & \SOUND                    & \SOUND                    & \SOUND                    & \SOUND                    & \PASS                     \\
            \ColHead{apache-commons-compress-cc-delta-07-vuln\_8}          & \SOUND                    & \SOUND                    & \SOUND                    & \SOUND                    & \SOUND                    & \PASS                     \\
            \ColHead{apache-commons-compress-cc-full-01-vuln\_0}           & \SOUND                    & \SOUND                    & \SOUND                    & \SOUND                    & \SOUND                    & \PASS                     \\
            \ColHead{apache-commons-compress-cc-full-01-vuln\_1}           & \SOUND                    & \SOUND                    & \SOUND                    & \SOUND                    & \SOUND                    & \PASS                     \\
            \ColHead{apache-commons-compress-cc-full-01-vuln\_2}           & \BAD                      & \BAD                      & \BAD                      & \BAD                      & \BAD                      & \FAIL                     \\
            \ColHead{apache-commons-compress-cc-full-01-vuln\_5}           & \SOUND                    & \SOUND                    & \SOUND                    & \SOUND                    & \SOUND                    & \PASS                     \\
            \ColHead{apache-poi-poi-delta-01-vuln\_5}                      & \SOUND                    & \SOUND                    & \SOUND                    & \SOUND                    & \SOUND                    & \PASS                     \\
            \ColHead{apache-poi-poi-delta-01-vuln\_6}                      & \BAD                      & \SOUND                    & \SOUND                    & \SOUND                    & \SOUND                    & \PASS                     \\
            \ColHead{apache-poi-poi-full-01-vuln\_0}                       & \BAD                      & \SOUND                    & \SOUND                    & \BAD                      & \BAD                      & \PASS                     \\
            \ColHead{apache-poi-poi-full-01-vuln\_1}                       & \SOUND                    & \SOUND                    & \SOUND                    & \SOUND                    & \SOUND                    & \PASS                     \\
            \ColHead{apache-poi-poi-full-01-vuln\_2}                       & \SOUND                    & \BAD                      & \SOUND                    & \SOUND                    & \SOUND                    & \PASS                     \\
            \ColHead{apache-poi-poi-full-01-vuln\_3}                       & \BAD                      & \BAD                      & \BAD                      & \BAD                      & \BAD                      & \FAIL                     \\
            \ColHead{apache-poi-poi-full-01-vuln\_4}                       & \BAD                      & \BAD                      & \BAD                      & \BAD                      & \BAD                      & \FAIL                     \\
            \ColHead{curl-cu-delta-01-curl-005}                            & \SOUND                    & \SOUND                    & \SOUND                    & \SOUND                    & \SOUND                    & \PASS                     \\
            \ColHead{curl-cu-delta-02-curl-006}                            & \SOUND                    & \SOUND                    & \SOUND                    & \SOUND                    & \SOUND                    & \PASS                     \\
            \ColHead{curl-cu-delta-03-curl-007}                            & \SOUND                    & \SOUND                    & \SOUND                    & \SOUND                    & \SOUND                    & \PASS                     \\
            \ColHead{curl-cu-delta-04-curl-003}                            & \BAD                      & \BAD                      & \BAD                      & \BAD                      & \SOUND                    & \PASS                     \\
            \ColHead{curl-cu-delta-04-curl-008}                            & \SOUND                    & \SOUND                    & \SOUND                    & \SOUND                    & \SOUND                    & \PASS                     \\
            \ColHead{curl-cu-delta-05-curl-001}                            & \BAD                      & \SOUND                    & \SOUND                    & \SOUND                    & \SOUND                    & \PASS                     \\
            \ColHead{curl-cu-delta-05-curl-002}                            & \SOUND                    & \SOUND                    & \SOUND                    & \SOUND                    & \SOUND                    & \PASS                     \\
            \ColHead{curl-cu-full-01-curl-004}                             & \SOUND                    & \SOUND                    & \SOUND                    & \SOUND                    & \SOUND                    & \PASS                     \\
            \ColHead{dav1d-da-full-01-dav1d-001}                           & \BAD                      & \BAD                      & \BAD                      & \BAD                      & \BAD                      & \FAIL                     \\
            \ColHead{freerdp-fp-delta-01-vuln\_001}                        & \SOUND                    & \SOUND                    & \SOUND                    & \SOUND                    & \SOUND                    & \PASS                     \\
            \ColHead{freerdp-fp-delta-02-vuln\_002}                        & \SOUND                    & \BAD                    & \SOUND                    & \SOUND                    & \SOUND                    & \PASS                     \\
            \ColHead{freerdp-fp-delta-03-vuln\_003}                        & \SOUND                    & \SOUND                    & \SOUND                    & \SOUND                    & \SOUND                    & \PASS                     \\
            \ColHead{freerdp-fp-full-01-vuln\_004}                         & \SOUND                    & \SOUND                    & \SOUND                    & \SOUND                    & \SOUND                    & \PASS                     \\
            \ColHead{freerdp-fp-full-01-vuln\_005}                         & \SOUND                    & \SOUND                    & \SOUND                    & \SOUND                    & \SOUND                    & \PASS                     \\
            \ColHead{lcms-cm-full-01-lcms-001}                             & \SOUND                    & \SOUND                    & \SOUND                    & \SOUND                    & \SOUND                    & \PASS                     \\
            \ColHead{lcms-cm-full-01-lcms-002}                             & \BAD                      & \SOUND                    & \SOUND                    & \SOUND                    & \SOUND                    & \PASS                     \\
            \ColHead{libavif-av-delta-02-avif-002}                         & \SOUND                    & \SOUND                    & \SOUND                    & \SOUND                    & \SOUND                    & \PASS                     \\
            \ColHead{libexif-ex-delta-01-exif-003}                         & \SOUND                    & \SOUND                    & \SOUND                    & \SOUND                    & \SOUND                    & \PASS                     \\
            \ColHead{libexif-ex-delta-01-exif-004}                         & \SOUND                    & \SOUND                    & \SOUND                    & \SOUND                    & \SOUND                    & \PASS                     \\
            \ColHead{libexif-ex-delta-01-exif-005}                         & \SOUND                    & \SOUND                    & \SOUND                    & \SOUND                    & \SOUND                    & \PASS                     \\
            \ColHead{libexif-ex-delta-02-exif-001}                         & \SOUND                    & \SOUND                    & \SOUND                    & \SOUND                    & \SOUND                    & \PASS                     \\
            \ColHead{libexif-ex-delta-03-exif-002}                         & \BAD                      & \SOUND                    & \SOUND                    & \SOUND                    & \SOUND                    & \PASS                     \\
            \ColHead{libxml2-lx-delta-01-vuln\_001}                        & \SOUND                    & \SOUND                    & \SOUND                    & \SOUND                    & \SOUND                    & \PASS                     \\
            \ColHead{libxml2-lx-delta-02-vuln\_002}                        & \SOUND                    & \SOUND                    & \SOUND                    & \SOUND                    & \SOUND                    & \PASS                     \\
            \ColHead{libxml2-lx-delta-03-vuln\_004}                        & \SOUND                    & \BAD                    & \SOUND                    & \SOUND                    & \SOUND                    & \PASS                     \\
            \ColHead{libxml2-lx-ex1-delta-01-vuln\_001}                    & \SOUND                    & \SOUND                    & \SOUND                    & \SOUND                    & \SOUND                    & \PASS                     \\
            \ColHead{libxml2-lx-full-01-vuln\_003}                         & \BAD                      & \SOUND                    & \SOUND                    & \SOUND                    & \SOUND                    & \PASS                     \\
            \ColHead{libxml2-lx-full-01-vuln\_005}                         & \BAD                      & \SOUND                    & \SOUND                    & \SOUND                    & \SOUND                    & \PASS                     \\
            \ColHead{log4j2-log4j2-delta-01-vuln\_0}                       & \SOUND                    & \SOUND                    & \SOUND                    & \SOUND                    & \SOUND                    & \PASS                     \\
            \ColHead{mongoose-mg-delta-01-mongoose\_1}                     & \BAD                      & \SOUND                    & \SOUND                    & \SOUND                    & \SOUND                    & \PASS                     \\
            \ColHead{mongoose-mg-delta-02-mongoose\_2}                     & \BAD                      & \BAD                      & \BAD                      & \BAD                      & \BAD                      & \FAIL                     \\
            \ColHead{mongoose-mg-full-01-mongoose\_0}                      & \BAD                      & \SOUND                    & \SOUND                    & \SOUND                    & \SOUND                    & \PASS                     \\
            \ColHead{pdfbox-pdfbox-delta-01-vuln\_2}                       & \SOUND                    & \SOUND                    & \SOUND                    & \SOUND                    & \SOUND                    & \PASS                     \\
            \ColHead{pdfbox-pdfbox-full-01-vuln\_0}                        & \SOUND                    & \BAD                    & \SOUND                    & \SOUND                    & \SOUND                    & \PASS                     \\
            \ColHead{pdfbox-pdfbox-full-01-vuln\_1}                        & \SOUND                    & \SOUND                    & \SOUND                    & \SOUND                    & \SOUND                    & \PASS                     \\
            \ColHead{pdfbox-pdfbox-full-01-vuln\_3}                        & \SOUND                    & \SOUND                    & \SOUND                    & \SOUND                    & \SOUND                    & \PASS                     \\
            \ColHead{pdfbox-pdfbox-full-01-vuln\_4}                        & \SOUND                    & \SOUND                    & \SOUND                    & \SOUND                    & \SOUND                    & \PASS                     \\
            \ColHead{pdfbox-pdfbox-full-01-vuln\_5}                        & \SOUND                    & \SOUND                    & \SOUND                    & \SOUND                    & \SOUND                    & \PASS                     \\
            \ColHead{pdfbox-pdfbox-full-01-vuln\_6}                        & \SOUND                    & \BAD                      & \SOUND                    & \SOUND                    & \SOUND                    & \PASS                     \\
            \ColHead{pdfbox-pdfbox-full-01-vuln\_7}                        & \SOUND                      & \BAD                      & \SOUND                    & \SOUND                    & \SOUND                    & \PASS                     \\
            \ColHead{pdfbox-pdfbox-full-01-vuln\_8}                        & \SOUND                    & \BAD                      & \BAD                      & \BAD                      & \SOUND                    & \PASS                     \\
            \ColHead{shadowsocks-shadowsocks-full-01-shadowsocks-libev\_0} & \SOUND                    & \SOUND                    & \SOUND                    & \SOUND                    & \SOUND                    & \PASS                     \\
            \ColHead{shadowsocks-shadowsocks-full-01-shadowsocks-libev\_1} & \SOUND                    & \SOUND                    & \SOUND                    & \SOUND                    & \SOUND                    & \PASS                     \\
            \ColHead{shadowsocks-shadowsocks-full-01-shadowsocks-libev\_2} & \SOUND                    & \SOUND                    & \SOUND                    & \SOUND                    & \SOUND                    & \PASS                     \\
            \ColHead{shadowsocks-shadowsocks-full-01-shadowsocks-libev\_3} & \SOUND                    & \SOUND                    & \SOUND                    & \SOUND                    & \SOUND                    & \PASS                     \\
            \ColHead{shadowsocks-shadowsocks-full-01-shadowsocks-libev\_4} & \SOUND                    & \SOUND                    & \SOUND                    & \SOUND                    & \SOUND                    & \PASS                     \\
            \ColHead{systemd-systemd-full-001-systemd-001}                 & \SOUND                    & \SOUND                    & \SOUND                    & \SOUND                    & \SOUND                    & \PASS                     \\
            \ColHead{systemd-systemd-full-001-systemd-003}                 & \SOUND                    & \SOUND                    & \SOUND                    & \SOUND                    & \SOUND                    & \PASS                     \\
            \ColHead{systemd-systemd-full-001-systemd-004}                 & \SOUND                    & \SOUND                    & \SOUND                    & \SOUND                    & \SOUND                    & \PASS                     \\
            \ColHead{systemd-systemd-full-001-systemd-005}                 & \BAD                      & \SOUND                    & \SOUND                    & \SOUND                    & \SOUND                    & \PASS                     \\
            \ColHead{tika-tk-delta-01-vuln\_9}                             & \BAD                      & \BAD                      & \BAD                      & \SOUND                    & \BAD                      & \PASS                     \\
            \ColHead{tika-tk-delta-02-vuln\_3}                             & \SOUND                    & \SOUND                    & \SOUND                    & \SOUND                    & \SOUND                    & \PASS                     \\
            \ColHead{tika-tk-delta-03-vuln\_4}                             & \SOUND                    & \SOUND                    & \SOUND                    & \SOUND                    & \SOUND                    & \PASS                     \\
            \ColHead{tika-tk-delta-04-vuln\_5}                             & \SOUND                    & \SOUND                    & \SOUND                    & \SOUND                    & \SOUND                    & \PASS                     \\
            \ColHead{tika-tk-delta-05-vuln\_2}                             & \BAD                      & \SOUND                    & \SOUND                    & \SOUND                    & \SOUND                    & \PASS                     \\
            \ColHead{tika-tk-delta-06-vuln\_10}                            & \BAD                      & \SOUND                    & \SOUND                    & \SOUND                    & \SOUND                    & \PASS                     \\
            \ColHead{tika-tk-full-01-vuln\_0}                              & \SOUND                    & \SOUND                    & \SOUND                    & \SOUND                    & \SOUND                    & \PASS                     \\
            \ColHead{tika-tk-full-01-vuln\_1}                              & \SOUND                    & \BAD                      & \SOUND                    & \SOUND                    & \BAD                      & \PASS                     \\
            \ColHead{tika-tk-full-01-vuln\_6}                              & \SOUND                    & \SOUND                    & \SOUND                    & \SOUND                    & \SOUND                    & \PASS                     \\
            \ColHead{tika-tk-full-01-vuln\_7}                              & \SOUND                    & \SOUND                    & \SOUND                    & \SOUND                    & \SOUND                    & \PASS                     \\
            \ColHead{tika-tk-full-01-vuln\_8}                              & \SOUND                    & \SOUND                    & \SOUND                    & \SOUND                    & \SOUND                    & \PASS                     \\
            \ColHead{wireshark-ws-delta-01-vuln\_003}                      & \SOUND                    & \BAD                      & \SOUND                    & \SOUND                    & \SOUND                    & \PASS                     \\
            \ColHead{wireshark-ws-delta-02-vuln\_004}                      & \BAD                      & \SOUND                    & \SOUND                    & \SOUND                    & \SOUND                    & \PASS                     \\
            \ColHead{wireshark-ws-delta-03-vuln\_006}                      & \SOUND                    & \SOUND                    & \SOUND                    & \SOUND                    & \SOUND                    & \PASS                     \\
            \ColHead{wireshark-ws-delta-04-vuln\_007}                      & \SOUND                    & \SOUND                    & \SOUND                    & \SOUND                    & \SOUND                    & \PASS                     \\
            \ColHead{wireshark-ws-delta-05-vuln\_008}                      & \SOUND                    & \SOUND                    & \SOUND                    & \SOUND                    & \SOUND                    & \PASS                     \\
            \ColHead{wireshark-ws-delta-07-vuln\_013}                      & \SOUND                    & \SOUND                    & \SOUND                    & \SOUND                    & \SOUND                    & \PASS                     \\
            \ColHead{wireshark-ws-full-01-vuln\_001}                       & \SOUND                    & \SOUND                    & \SOUND                    & \SOUND                    & \SOUND                    & \PASS                     \\
            \ColHead{wireshark-ws-full-01-vuln\_002}                       & \SOUND                    & \SOUND                    & \SOUND                    & \SOUND                    & \SOUND                    & \PASS                     \\
            \ColHead{wireshark-ws-full-01-vuln\_005}                       & \SOUND                    & \SOUND                    & \SOUND                    & \SOUND                    & \SOUND                    & \PASS                     \\
            \ColHead{wireshark-ws-full-01-vuln\_010}                       & \SOUND                    & \SOUND                    & \SOUND                    & \SOUND                    & \SOUND                    & \PASS                     \\
            \ColHead{wireshark-ws-full-01-vuln\_011}                       & \SOUND                    & \SOUND                    & \SOUND                    & \SOUND                    & \SOUND                    & \PASS                     \\
            \ColHead{wireshark-ws-full-01-vuln\_012}                       & \SOUND                    & \SOUND                    & \SOUND                    & \SOUND                    & \SOUND                    & \PASS                     \\
            \ColHead{xz-xz-full-01-xz-001}                                 & \SOUND                    & \SOUND                    & \SOUND                    & \SOUND                    & \SOUND                    & \PASS                     \\
            \hline
            \RowHead{Totals}                                               & \RowHead{72/92\n(78.3\%)} & \RowHead{76/92\n(82.6\%)} & \RowHead{84/92\n(91.3\%)} & \RowHead{84/92\n(91.3\%)} & \RowHead{83/92\n(90.2\%)} & \RowHead{87/92\n(94.6\%)} \\
            \hline
        \end{tabular}
    }
    \label{tab:vuln-eval}
\end{table*}

\end{document}